\documentclass[
reprint,
 amsmath,amssymb,
aps,
superscriptaddress
]{revtex4-2}

\usepackage{graphicx}
\usepackage{dcolumn}
\usepackage{bm}%
\usepackage{xcolor}

\begin{document}

\title{Terahertz transitions in finite carbon chains}

\author{R. R. Hartmann}
\email{richard.hartmann@dlsu.edu.ph}
\affiliation{Physics Department, De La Salle University,  2401 Taft Avenue, 0922 Manila, Philippines}

\author{S. Kutrovskaya}
\affiliation{School of Science, Westlake University, 18 Shilongshan Road, Hangzhou 310024, Zhejiang Province, China}
\affiliation{Institute of Natural Sciences, Westlake Institute for Advanced Study, 18 Shilongshan Road, Hangzhou 310024, Zhejiang Province, China}
\affiliation{Department of Physics and Applied Mathematics, Stoletov Vladimir State University, 600000 Gor’kii street, Vladimir, Russia}

\author{A. Kucherik}
\affiliation{Department of Physics and Applied Mathematics, Stoletov Vladimir State University, 600000 Gor’kii street, Vladimir, Russia}

\author{A. V. Kavokin}
\affiliation{School of Science, Westlake University, 18 Shilongshan Road, Hangzhou 310024, Zhejiang Province, China}
\affiliation{Institute of Natural Sciences, Westlake Institute for Advanced Study, 18 Shilongshan Road, Hangzhou 310024, Zhejiang Province, China}
\affiliation{Physics and Astronomy, University of Southampton, Highfield, Southampton, SO17 1BJ, United Kingdom}
\affiliation{Russian Quantum Center, Skolkovo IC, Bolshoy Bulvar 30, bld. 1, Moscow 121205, Russia}
\affiliation{NTI Center for Quantum Communications, National University of Science and Technology MISiS, Moscow 119049,
Russia}

\author{M. E. Portnoi}
\email{M.E.Portnoi@exeter.ac.uk}
\affiliation{Physics and Astronomy, University of Exeter, Stocker Road, Exeter EX4 4QL, United Kingdom}
\affiliation{ITMO University, St.Petersburg 197101, Russia}

\begin{abstract}
We predict an optical effect associated with systems which exhibit topologically protected states separated by a finite distance. We develop a tight-binding model to calculate the optical selection rules in linear chains of atoms of
different lengths, and show the crucial importance of edge states. For long enough molecules the
interband transitions involving these edge states are in the highly sought-after THz frequency range. Although we have specifically considered finite carbon chains terminated by gold nanoparticles, the main results of our work can be generalized to various systems which exhibit topologically protected states separated by a finite distance.
\end{abstract}

\maketitle

\section{Introduction}
Carbyne has attracted much interest and its fair share of controversy since the first claims of its discovery in the 1960s~\cite{kasatoch1967crystalline,el1968new,whittaker1969carbon,sladkov1969diamond,smith1981graphitic,smith1982carbyne}, a detailed history of which can be found in Ref.~\cite{heimann1999carbyne}. 
The strong historic opposition against the existence of carbynes as a stable allotrope of carbon is, in part, associated with the high reactivity of carbon double and triple bonds, and the prediction that one-dimensional crystals would be thermodynamically unstable and therefore not able to exist~\cite{lifshitz2013statistical}. Both of the aforementioned reasons make the the synthesis of carbyne chains a challenging endeavor. Various synthesis methods have been explored such as using end-cap groups for stabilization~\cite{johnson1972silylation,gibtner2002end,chalifoux2010synthesis}, encapsulating polyyne molecules in carbon nanotubes~\cite{nishide2006single,shi2016confined}, and synthesizing polyynes in a submerged electric arc in organic solvents~\cite{cataldo2004synthesis}. Carbynes of finite length have also been synthesized using a novel laser ablation in liquid~\cite{pan2015carbyne,kucherik2016two}, and recent advances using this method have demonstrated the synthesis of stable elongated carbon chains up to 24 atoms long~\cite{kutrovskaya2020excitonic}. The mechanical stabilization of carbyne in this instance was achieved due to the electron bonding of the carbon chains with gold nanoparticles, inhibiting the vibration-induced decomposition into shorter components, folding and bending. In solution, the Peierls deformation \cite{bianco2018carbon} is compensated by the viscosity of the surrounding medium and the attraction to the metal anchors. These chains can then be deposited onto a solid substrate~\cite{kutrovskaya2020excitonic}.

\begin{figure}[h]
    \centering
    \includegraphics[width=0.7\linewidth]{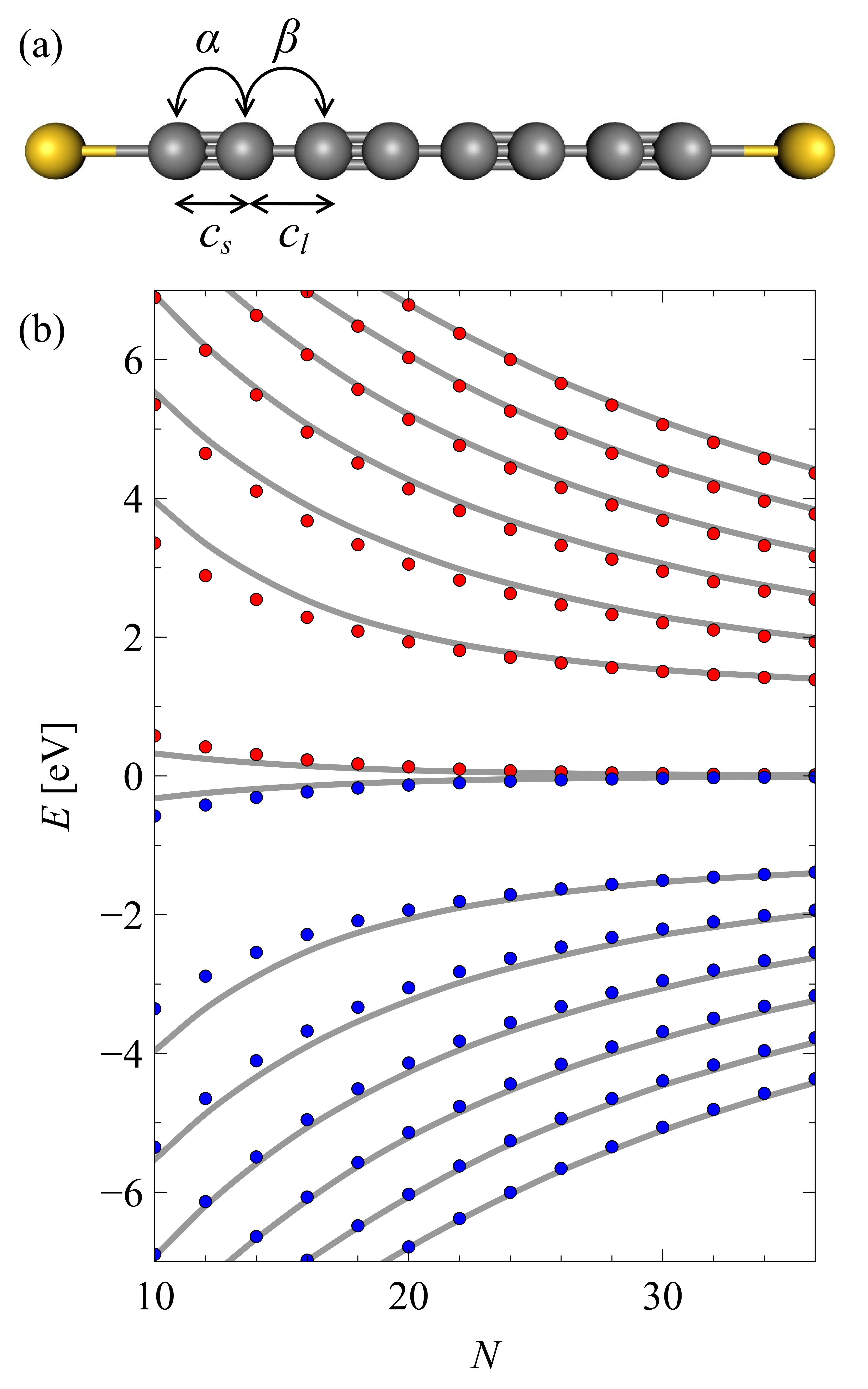}
    \caption{(a) Schematic representation of a polyyne carbyne chain with alternating single and triple bonds, terminated by two gold atoms. The grey and yellows spheres depict the carbon and gold atoms respectively, and the single and triple bonds lengths are given by $c_l$ and $c_s$, while $\alpha$ and $\beta$ are their corresponding tight-binding parameters. (b) The 7 highest occupied and lowest unoccupied molecular energy levels, $E$, of a carbyne chain of various lengths, defined by the tight-binding parameters $\alpha=-4.657$~eV, $\beta=-3.548$~eV and $\Delta_1=\Delta_2=0$. Here $N$ denotes the number of atoms in the chain and the gray lines are the approximate eigenvalue expressions.
    }
    \label{fig:chain_schematic}
\end{figure}

Although there is still some contention whether or not ``pure'' carbyne can truly exist, part of the controversy arguably derives from its ambiguous definition. Ignoring the presence of any end groups, carbyne comes in two isomeric forms, a conjugated triply bonded form (polyyne) and a cumulated doubly bonded form (polycumulene). In this paper, we consider finite polyyne chains, i.e., a series of consecutive alkynes $\left(-\mathrm{C}\equiv\mathrm{C}-\right)_{n}$ with $n$ greater than 1, with the end carbon atoms not necessarily being terminated by hydrogen atoms, but for example attached to gold nanoparticles (see Fig.~\ref{fig:chain_schematic} (a)). For these chains we study optical transitions which arise due to finite size effects. Unlike the case of an infinite polyyne chain, finite chains possess edge states. In the infinite length limit, the two, clearly defined states (one located at each edge) would be degenerate in energy. However, for finite chains the edge-state wavefunctions overlap to form a bonding and anti-bonding state. The splitting between these symmetric and anti-symmetric energy states is simply related to the overlap of the two wavefunctions (just like in a Dirac double well~\cite{hartmann2020guided}, or a bipolar waveguide~\cite{hartmann2020bipolar}). For long enough chains the overlap is sufficiently weak to generate a THz gap. Naturally, dipole transitions between them are allowed as the states are of opposite parity, and thus polyyne chains present a new avenue for the generation of THz radiation.

In contrast to graphite, graphene and carbon nanotubes, carbynes exhibit strong optical emission. Indeed, unlike graphite or graphene which are strong absorbers and do not emit light, and carbon nanotubes where the multi-valley band structure leads to dark excitons suppressing the luminescence, carbyne is often called ``white carbon" because of its ability to emit visible light.  Carbon chains exhibit a strong dependence of their optical band gap on chain length~\cite{al2014electronic,pan2015carbyne}. Although the presence of the gap in the energy spectrum of an infinite carbyne chain is well
known~\cite{heimann1999carbyne}, what is less well known is the emergence of two edge states within the gap when the chain is of finite size. 
It should be noted that although structural distortions like solitons and polarons also lead to extra levels appearing in the optical gap~\cite{rice1986solitons}, these are often associated with chains charged via doping with strong electron withdrawing species, which is beyond the scope of this paper, and a topic for future study.

Simple, tractable tight-binding models have been successfully employed to explain both the electronic and optical properties of various low-dimensional forms of carbon~\cite{dresselhaus1998physical,Hartmann_2014_Rev}, providing spectacular success in explaining the main electronic properties of graphene~\cite{neto2009electronic}. For example, by using a simple zone-folding model, the approximate band structure of a carbon nanotube can be obtained from the band structure of graphene along allowed lines in k-space defined by the chiral vector~\cite{dresselhaus1998physical}. However, applying a similar approach to derive the spectrum of a finite carbyne chain from an infinite one by simply applying the periodic boundary condition will lead to an incomplete spectrum. Namely, the two edge states, which correspond to the highest occupied molecular orbital (HOMO) and lowest unoccupied molecular orbital (LUMO) are absent. An alternative approach is to calculate the full energy spectrum from a finite matrix approach.
Recent advances in the study of tridiagonal matrices~\cite{losonczi1992eigenvalues,yueh2005eigenvalues,kouachi2006eigenvalues,da2007characteristic,willms2008analytic,kouachi2008eigenvalues,da2019eigenpairs} have opened the door to the quasi-analytical description of the optical properties of finite-length carbyne chains via the tight-binding approximation. In what follows we calculate the complete eigenvalue spectrum of carbyne chains terminated by gold atoms, and use the analytic eigenvectors to calculate the optical selection rules. We demonstrate the presence of two edge states, derive a condition for their existence, and show that optical transitions between them are allowed. For long enough molecules they occur in the highly desirable terahertz (THz) frequency range. 


\section*{Theoretical Model}
We shall now present a simple, analytically tractable model, which can be used to understand and explain the optical transitions between the electronic states of finite polyyne chains. We adopt the same nearest-neighbor tight-binding model which is commonly used in carbon-based materials~\cite{dresselhaus1998physical}. This model yielded spectacular success in describing their electronic and optical properties, especially in the case of carbon nanotubes and graphene~\cite{Hartmann_2014_Rev}. Using the nearest neighbor tight-binding model~\cite{dresselhaus1998physical}, the Hamiltonian of an infinite polyyne chain can be written as
\begin{equation}
\hat{H}=\left[\begin{array}{cc}
0 & f^{\star}\\
f & 0
\end{array}\right],
\label{eq:Ham_con}
\end{equation}
where $f=\alpha\exp\left(ik c_{s}\right)+\beta\exp\left(-ik c_{l}\right)$, $k$ is the wave vector along the chain, $c_{s}$ and $c_{l}$ are the distance between neighboring carbon atoms associated with the short (triple) and long (single) covalent bonds, and their corresponding transfer integrals are $\alpha$ and $\beta$, respectively. 
The Hamiltonian given in Eq.~(\ref{eq:Ham_con}) acts on a two-component wavefunction, with each component associated with one of the two sublattices of the chain. The energy spectrum of a particle described by the Hamiltonian Eq.~(\ref{eq:Ham_con}) is 
\begin{equation}
E=\pm\sqrt{\alpha^{2}+\beta^{2}+2\alpha\beta\cos\left(k c\right)},
\label{eq:infinite_energy}
\end{equation}
where 
$c=c_s+c_l$ is the lattice constant, and $E_{g}$, the band gap, is given by the expression $2\left|\alpha-\beta\right|$. However, there is some controversy surrounding the precise value of the band gap, since various numerical methods yield drastically different results~\cite{deretzis2011coherent,al2014electronic}, with a range of almost 4~eV between them. It should be noted that the approximate, albeit incomplete, band structure of a finite-length polyyne chain can be obtained by using a simple zone-folding approximation of the band structure of the infinite chain. Applying the Born--von Karman boundary condition to the infinite chain leads to discretized energy levels, with corresponding quantised wavenumbers $k=\pi n/L$, where $L$ is the length of the chain. However, applying this condition does not give rise to the emergence of an edge state. It should also be noted that inspection of 
Eq.~(\ref{eq:infinite_energy}) reveals no quantitative difference between the case of $\alpha>\beta$ and $\alpha<\beta$.

In what follows we shall use the nearest-neighbour tight-binding approximation to study chains of finite lengths. As mentioned in the introduction, the HOMO-LUMO gap is strongly dependent on the length of the chain. An important feature of our finite chains is the presence of edge states, which are topologically protected like in certain types of graphene nanoribbons~\cite{PhysRevB.54.17954} and semiconductor superlattices~\cite{STESLICKA200293}. Moving from an infinite chain to a finite chain results in changes to the energy spectrum given in Eq.~(\ref{eq:infinite_energy}). The quantisation of momentum discretizes the energy spectrum, while the presence of edges results in two energy levels appearing within what would have been the band gap if the chain were infinite. Notably, these two levels are the HOMO and LUMO levels of undoped chains. The Hamiltonian of a finite polyyne chain, schematically depicted in Fig. \ref{fig:chain_schematic}~(a), composed of $N-2$ carbon atoms (where $N$ is an even integer) and a differing atom on each end, can be written as:
\begin{equation}
\mathbf{H}=
\left(\begin{array}{ccccccc}
-\Delta_{1} & b_{1}\\
b_{1} & 0 & b_{2}\\
 & b_{2} & \ddots & \ddots\\
 &  & \ddots & 0 & b_{j}\\
 &  &  & b_{j} & \ddots & \ddots\\
 &  &  &  & \ddots & 0 & b_{N-1}\\
 &  &  &  &  & b_{N-1} & -\Delta_{2}
\end{array}\right),
\label{eq:Ham_finite}
\end{equation}
where
\begin{equation}
b_{j}=\left\{ \begin{array}{c}
\beta,\,\mathrm{if} j \mathrm{\,is\,odd}\\
\alpha,\,\mathrm{if} j \mathrm{\,is\,even}
\end{array}\right.,\qquad j=1,\,2,\,\ldots,N,
\nonumber
\end{equation}
and $-\Delta_{1}$ and $-\Delta_{2}$ are the on-site energies of the edge atoms measured with respect to the on-site energies of the carbon atoms. When $N=2m$ is even, where $m$ is a positive integer, the characteristic polynomial of $\mathbf{H}-E\mathbf{I}$, is given by
\begin{equation}
\begin{split}
\mathrm{det}
\left(\mathbf{H}-E\mathbf{I}\right)=\left(\alpha\beta\right)^{m-1}
\bigg\{
\alpha\beta U_{m}+\left[\Delta_{1}\Delta_{2}+\alpha^{2}
\right.
\bigg.
\\
+\bigg.\left.
\left(\Delta_{1}+\Delta_{2}\right)E\right]U_{m-1}+\frac{\beta\Delta_{1}\Delta_{2}}{\alpha}U_{m-2}
\bigg\},
\end{split}
\label{eq:secular}
\end{equation}
where $U_{m}=U_{m}\left(\cos\left(\theta\right)\right)$ are the Chebyshev polynomials of the second kind and 
\begin{equation}
\cos\left(\theta\right)=\frac{E^{2}-\alpha^{2}-\beta^{2}}{2\alpha\beta}.
\label{eq:angel}
\end{equation}
In general, Eq.~(\ref{eq:secular}) must be solved numerically. However, within the long chain limit, i.e. $N\gg1$, approximate expressions for the eigenvalues can be obtained. For all chain lengths the eigenvectors can be expressed exactly in terms of the numerically or analytically obtained eigenvalues. We denote the energy eigenvalues, which satisfy the condition $\mathrm{det}\left(\textbf{H}-E\textbf{I}\right)=0$, as $E_{p}$, where $p=1,\,2,\,\ldots,\,N$. We use the convention that places the eigenvalues in ascending order, with $E_{1}$ being the lowest energy level, and $E_{N}$ being the highest one. The non-normalized eigenvector, $\mathrm{\boldsymbol{u}}^{p}=\left(u_{1}^{p},\dots,u_{j}^{p},\dots,u_{N}^{p}\right)^{\mathrm{T}}$ belonging to the $p^{\mathrm{th}}$ eigenvalue, $E_{p}$, is given by the expressions \cite{kouachi2006eigenvalues}:
\begin{equation}
\begin{split}
u_{j}^{p}&=C_{p}\left[\alpha\left(E_p+\Delta_{2}\right)\sin\left(\frac{N-j+1}{2}\theta_{p}\right)
\right.
\\
&+
\left.
\beta\Delta_{2}\sin\left(\frac{N-j-1}{2}\theta_{p}\right)
\right]
\end{split}
\label{evector_1}
\end{equation}
for odd $j$ and 
\begin{equation}
\begin{split}
u_{j}^{p}&=C_{p}
\left[\alpha\beta\sin\left(\frac{N-j+2}{2}\theta_{p}\right)
\right.
\\
&+
\left.
\left(\alpha^{2}
+E_p\Delta_{2}\right)\sin\left(\frac{N-j}{2}\theta_{p}\right)
\right]
\end{split}
\label{evector_2}
\end{equation}
for even $j$, where $C_{p}$ is a normalisation constant, and $\theta_p$ is obtained from Eq.~(\ref{eq:angel}) for a particular $E_{p}$.

We shall now consider two energy regimes: First, when $\left|\left(E^{2}-\alpha^{2}-\beta^{2}\right)/2\alpha\beta\right|<1$, which corresponds to the case where $\theta$ is real, i.e., momentum is quantized like a particle in a box, and second, when $E^{2}<\left(\alpha-\beta\right)^{2}$, i.e. when $\theta$ is complex, in this instance the wavevector is imaginary, and we refer to these states as edge states. 
Let us first consider the case when $\theta$ is real, and $\Delta_{1}=\Delta_{2}=0$. In the infinite chain limit, i.e., $m\rightarrow\infty$, the solutions are simply $\theta=n\pi/m$, where $n=1,\,2,\,\ldots,\,m-1$. Therefore, for large chains we may expand Eq.~(\ref{eq:secular}) about the point $\theta=n\pi/m$, and obtain to a first-order approximation the eigenvalues
\begin{equation}
\begin{split}
E_{n}&=\pm\sqrt{\alpha^{2}+\beta^{2}+2\alpha\beta\cos\left(\theta_{n}\right)},\\
\theta_{n}&=\frac{n\pi}{m}-\frac{\beta\sin\left(\frac{n\pi}{m}\right)}{m\alpha+\beta\left(m+1\right)\cos\left(\frac{n\pi}{m}\right)}.
\end{split}
\label{eq:energy_wave}   
\end{equation}
Let us now consider the case when $\theta$ takes on the value of a complex number. In this instance we invoke the change of variable $\theta=\pi-i\varphi$. For the case of $\Delta_{1}=\Delta_{2}=0$, the secular equation, Eq.~(\ref{eq:secular}), results in
\begin{equation}
\frac{\sinh\left[\left(m+1\right)\varphi\right]}{\sinh\left(m\varphi\right)}-\frac{\alpha}{\beta}=0.
\label{eq:imag_angle}
\end{equation}
Since the minimum value of the function $\sinh\left[\left(m+1\right)\varphi\right]/\sinh\left(m\varphi\right)$ is $\left(m+1\right)/m$, for a solution to exist we require
\begin{equation}
    \frac{m+1}{m}-\frac{\alpha}{\beta}<0, 
\end{equation}
i.e., $\left|\beta\right|<\left|\alpha\right|$ and the chain has a minimum length to observe an edge state. A similar result can be obtained for non-zero values of $\Delta_{1}$ and $\Delta_{2}$:
\begin{equation}
\begin{split}
\frac{\left(m+1\right)}{m}-\frac{\alpha}{\beta}+\frac{\left(m-1\right)\Delta_{1}\Delta_{2}}{m\alpha^{2}}
\\-\frac{s\left(\Delta_{1}+\Delta_{2}\right)\left|\alpha-\beta\right|+\Delta_{1}\Delta_{2}}{\alpha\beta}<0
\end{split}
\end{equation}
It should be noted that the Su-Schrieffer-Heeger model~\cite{PhysRevLett.42.1698} can also be employed to show that the relative size of the hopping parameters dictates whether the edge states of a bipartite 1D lattice are topologically protected as the chain length increases (i.e., the winding number
differentiates between the cases of $\alpha>\beta$ and $\alpha<\beta$). This effect is well known and manifests itself in various physical systems, from semiconductor superlattices~\cite{Vladimirova1998} to optical cavity chains~\cite{PhysRevLett.116.046402,st2017lasing}.
This is in stark contrast to the result obtained by simply applying the Born--von Karman periodic boundary condition to an infinite chain, which will not yield an edge state. In the limit where $m\rightarrow\infty$, the solution of Eq.~(\ref{eq:imag_angle}) is $\varphi=\ln\left(\alpha/\beta\right)$, which corresponds to two edge states of zero energy. Therefore, for long chains we expand Eq.~(\ref{eq:imag_angle}) about the point $\varphi=\ln\left(\alpha/\beta\right)$ 
and to a first-order approximation we obtain the solution 
\begin{equation}
\varphi=\ln\left(\frac{\alpha}{\beta}\right)+\left[1-\left(\frac{\beta}{\alpha}\right)^{2}\right]\left(\frac{\beta}{\alpha}\right)^{N},
\label{eq:energy_phi}
\end{equation}
which yields the eigenvalues 
\begin{equation}
E=\pm\alpha\left(1-\frac{\beta^{2}}{\alpha^{2}}\right)\left(\frac{\beta}{\alpha}\right)^{m},
\label{eq:energy_edge}
\end{equation}
where the $-$ and $+$ signs correspond to the LUMO and HOMO  states respectively. 
Within the same regime, Eqs.~(\ref{eq:energy_phi}-\ref{eq:energy_edge}) allow the edge state wavefunctions to be written exactly:
\begin{equation}
\begin{split}
u_{j}^{p_e}&= 
\frac{i}{\sqrt{2}}\left[\frac{1-\left(\frac{\beta}{\alpha}\right)^{2}}{\left(\frac{\beta}{\alpha}\right)^{^{-N}}-1}\right]^{\frac{1}{2}}
\left[\left(\frac{\beta}{\alpha}\right)^{-\frac{j}{2}}\cos\left(\frac{j\pi}{2}\right)
\right.
\\
&+\left(-1\right)^{p_e+1}
\left.
\left(\frac{\beta}{\alpha}\right)^{-\frac{N-j+1}{2}}\sin\left(\frac{j\pi}{2}\right)\right],
\end{split}
\label{eq:exact_edge_wave}
\end{equation}
where $p_e=\left(N+1\mp1\right)/2$.
These edge state functions are strongly localized to the ends of the chain, 
whereas the other $N-2$ modes behave similarly to a one-dimensional particle in a box.

It can be clearly seen from Eq.~(\ref{eq:energy_edge}) that as the length of the chain increases, the difference in energy between the two edge-state levels becomes smaller and smaller, approaching zero in the limit of $m$ tending to infinity. Using the tight-binding parameters of Reference~\cite{al2014electronic}, a 36 atom long chain would yield a HOMO-LUMO gap of 
7.08~THz. In Fig.~\ref{fig:chain_schematic}~(b) we plot the energy levels obtained from Eq.~(\ref{eq:secular}) for chains of various lengths, using the values $\alpha=-4.657$~eV, $\beta=-3.548$~eV and $\Delta_1=\Delta_2=0$~eV~\cite{al2014electronic}. The gray lines depict the approximate expressions, Eqs.~(\ref{eq:energy_wave}) and~(\ref{eq:energy_edge}). It should be noted that when $\Delta_1=\Delta_2$, the eigenfunctions are purely even or odd, and within our eigenvalue numeration scheme the eigenfunctions of successive levels differ in parity. Thus, one edge state is symmetric, the other anti-symmetric, and as the chain length goes to infinity they form a degenerate ground state. Repeating the same process for the case of $\Delta^{2}\ll1$ allows the edge-state eigenvalues to be approximated by the expression:
\begin{equation}
\begin{split}
E&=s_{2}\alpha\left(1-\frac{\beta^{2}}{\alpha^{2}}\right)
\Bigg\{ \left(\frac{\beta}{\alpha}\right)^{2m}+\frac{\Delta_{1}^{2}+\Delta_{2}^{2}}{2\alpha^{2}}
\Bigg.
\\
&
\Bigg.
\pm\frac{\Delta_{1}+\Delta_{2}}{\alpha}\left[\left(\frac{\beta}{\alpha}\right)^{2m}+\left(\frac{\Delta_{1}-\Delta_{2}}{2\alpha}\right)^{2}\right]^{\frac{1}{2}}
\Bigg\} ^{\frac{1}{2}},
\end{split}
\end{equation}
where for the case of $\mathrm{sgn}\left(\pm\Delta_{1}\right)=1$, $s_{2}=\mathrm{sgn}\left(\Delta_{2}\pm\alpha\left(\beta/\alpha\right)^{m}\sqrt{\left|\Delta_{2}/\Delta_{1}\right|}\right)$, while when $\mathrm{sgn}\left(\pm\Delta_{1}\right)=-1$, $s_{2}=\mathrm{sgn}\left(\Delta_{1}\pm\alpha\left(\beta/\alpha\right)^{m}\sqrt{\left|\Delta_{2}/\Delta_{1}\right|}\right)$.


\section*{Optical transitions in finite chains}
For conjugated $\pi$-electron systems, the probability of an optical transition is
\begin{equation}
W_{i\rightarrow f}\propto\sum_{i,\,f}\left|\left\langle \psi_{i}
\left|\hat{\mathrm{e}}
\cdot{
\hat{\mathrm{v}}
}\right|
\psi_{f}\right\rangle \right|^{2}\delta\left(E_{f}-E_{i}-\hbar\omega\right),
\label{eq:Intensity}
\end{equation}
where $i$ and $f$ are the initial and final states, $E_{i}$ and $E_{f}$ are their corresponding energies and $\hat{\mathrm{e}}$ is the polarization unit vector of the electromagnetic wave. The velocity commutator is given by $\hat{\boldsymbol{\mathrm{v}}}=i\left[\mathbf{H}_{\mathrm{TB}},\,\mathbf{r}\right]$, where $\mathbf{H}_{\mathrm{TB}}$ is the tight-binding Hamiltonian of the system~\cite{dresselhaus1998physical}, $\mathbf{r}$ the position operator, and $\psi_{i}$ and $\psi_{f}$ the wavefunctions of the initial and final states, which are given by the linear combinations $\psi_{f}=\sum_{j}u_{j}^{f}\left|\phi_{j}\right\rangle$  and $\psi_{i}=\sum_{q}u_{q}^{i}\left|\phi_{q}\right\rangle$ , where $\left|\phi\right\rangle$ are the atomic orbital functions, and the weighting coefficients, $u_{j}^{f}$ and $u_{q}^{i}$, are given by Eqs.~(\ref{evector_1}) and~(\ref{evector_2}). Within the tight-binding approximation $\left\langle \phi_{j}\left|\mathbf{r}\right|\phi_{q}\right\rangle =\mathbf{r}_{j}\delta_{jq}$, where $\mathbf{r}_{j}$ is the position vector of the $j^{th}$ atom, and $\left\langle \phi_{j}\left|\mathbf{H}_{\mathrm{TB}}\right|\phi_{q}\right\rangle =H_{j,q}$, which are the matrix elements of Eq.~(\ref{eq:Ham_finite}). 
The matrix element of velocity along the chain can therefore be written as 
\begin{equation}
\begin{split}
\left|\left\langle \psi_{f}\left|\hat{\boldsymbol{\mathrm{v}}}\right|\psi_{i}\right\rangle \right|=
\left|\frac{1}{\hbar}\sum_{j=1}^{N}\left[\left(u_{j}^{f}\right)^{\star}u_{j+1}^{i}-\left(u_{j+1}^{f}\right)^{\star}u_{j}^{i}\right]\boldsymbol{l}_{j}H_{j+1,\,j}\right|,
\end{split}
\label{eq:VME}
\end{equation}
where $\boldsymbol{l}_j=\boldsymbol{r}_{j+1}-\boldsymbol{r}_{j}$ are the nearest-neighbor vectors connecting atoms $j$ and $j+1$. Although the tight-binding Hamiltonian only depends on the magnitude of the interatomic distances between nearest-neighbors, the exact geometric configuration of the molecule does enter into the matrix element of optical transition. In what follows we assume that our finite chains are straight due to the metallic clusters attached to their ends~\cite{kutrovskaya2020excitonic}. Furthermore, due to a lack of consensus on the exact value of the band-gap of an infinite chain we use the first order approximation $c_{s} = c_{l}$. However, the effect of bond length alternation can be fully accommodated by setting $\left|\boldsymbol{l}_{j}\right|=c/2$ and introducing the effective tight-binding parameters $\alpha\rightarrow\left[1-\left(\delta c/c\right)\right]\alpha$ and $\beta\rightarrow\left[1+\left(\delta c/c\right)\right]\beta$, where $\delta c=c_{l}-c_{s}$. 
It should be noted that the studied transitions are strongly polarized (anisotropic) and, as follows from Eq.~(\ref{eq:Intensity}), the intensity of all the considered transitions is proportional to squared cosine of the angle between the light polarization vector and the considered polyyne chain. This effect will be essential for experiments with arrays of carbon chains aligned in a regular manner on a solid substrate, in which case both optical absorption and THz emission will be strongly polarized. This is indeed similar to certain transitions in carbon nanotubes.

\begin{figure}[h]
    \centering
    \includegraphics[width=0.75\linewidth]{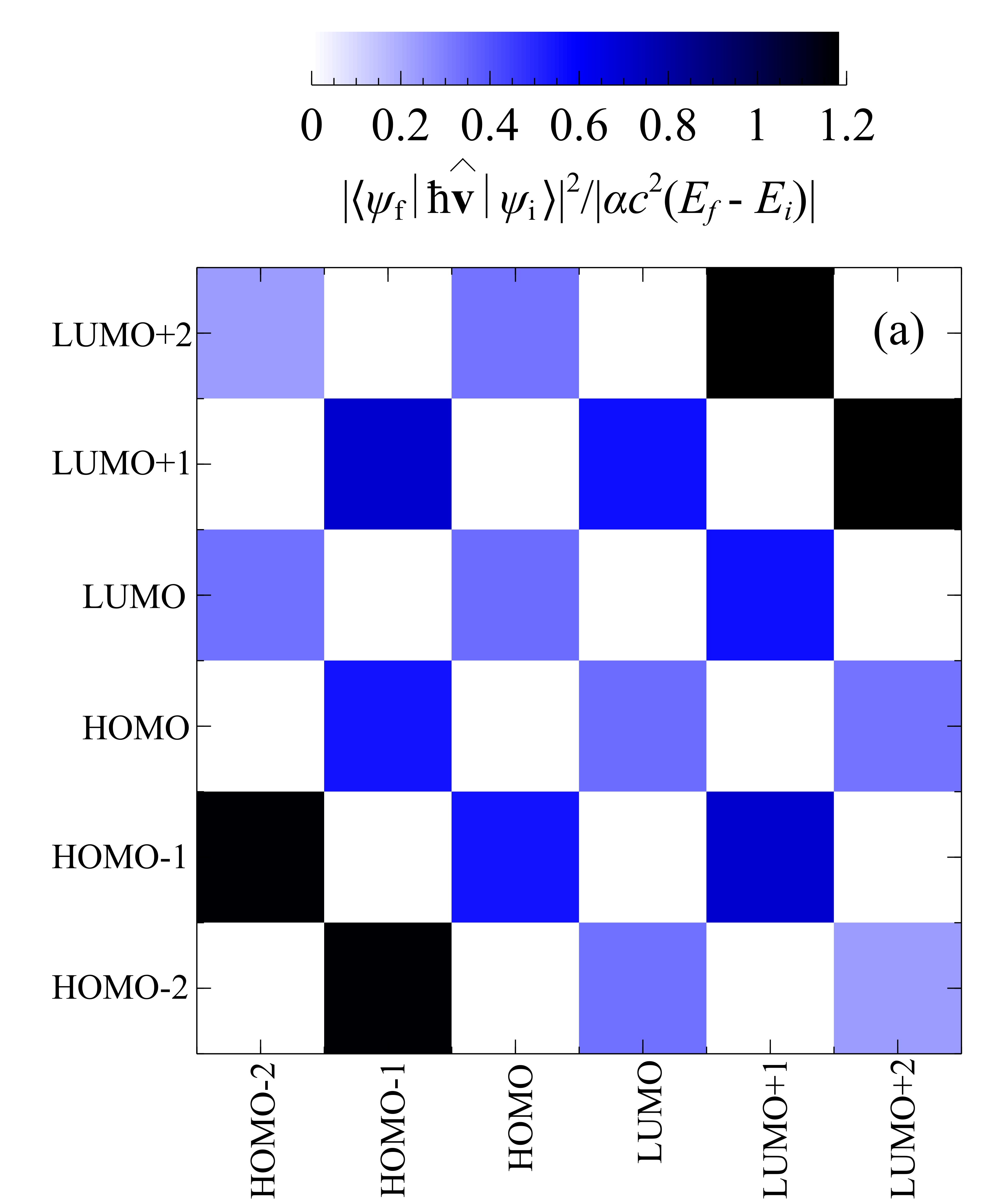}\\
    \includegraphics[width=0.75\linewidth]{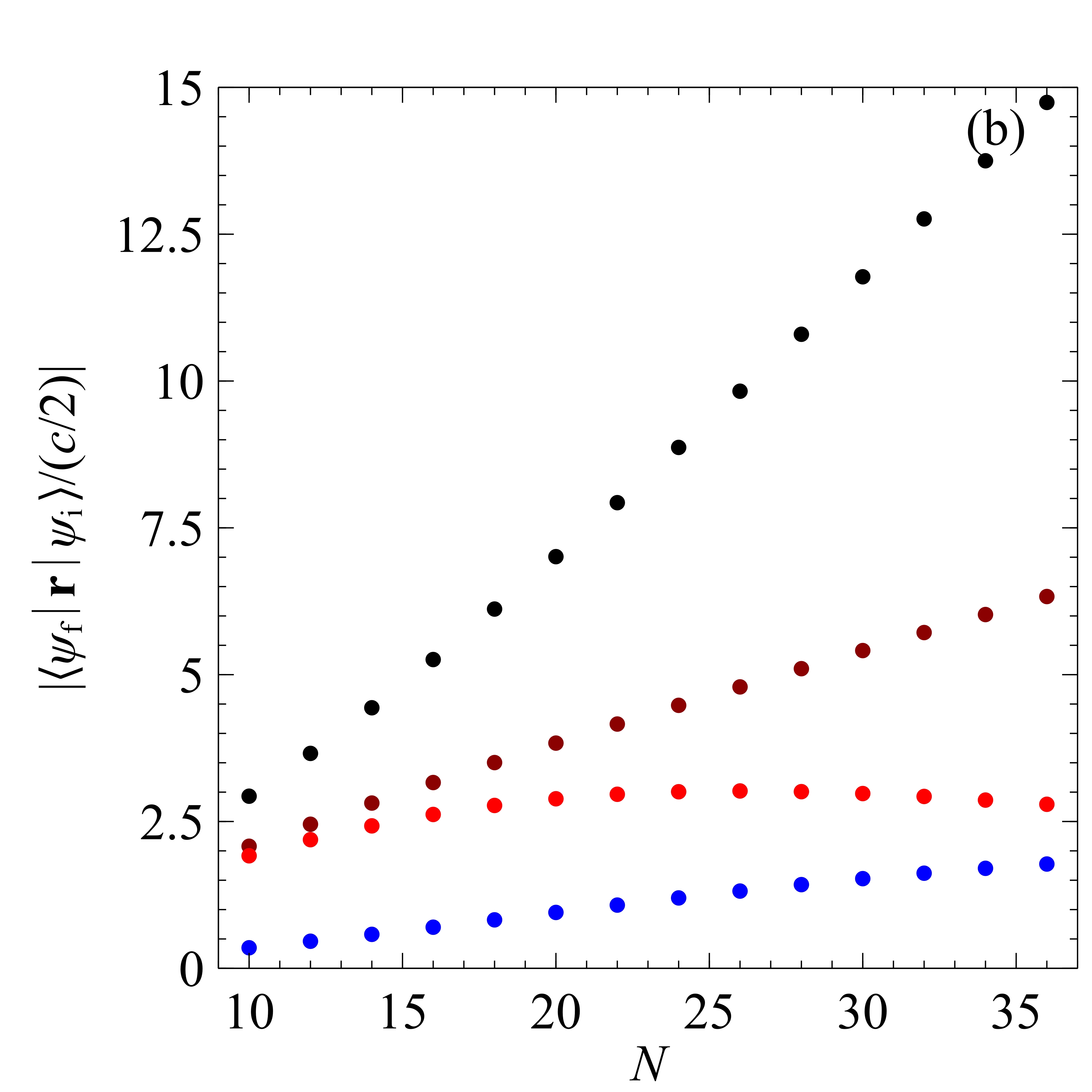}
    \caption{(a) 
   The absolute value of the oscillator strength, normalized by a dimensionless constant,} for transitions between energy levels of a chain comprised of $N=36$ atoms, with the lattice period $c$, tight-binding parameters $\alpha=-4.657$~eV, $\beta=-3.548$~eV, and the onsite energies of the edge atoms $\Delta_1=\Delta_2=-0.1$~eV. (b) The transition dipole moment, $\left|\left\langle \psi_{f}\left|\boldsymbol{r}\right|\psi_{i}\right\rangle /(c/2) \right|$, between: the LUMO and HOMO level (topmost black curve), LUMO+2 and LUMO+1 level (second from top maroon curve), LUMO+1 and LUMO level (third from top red curve), and lastly the HOMO and LUMO+2 level (bottommost blue curve), as a function of chain length, $N$, for the same tight-binding parameters.
    \label{figure_v_mat}
\end{figure}

\begin{figure}[h]
    \centering
    \includegraphics[width=0.9\linewidth]{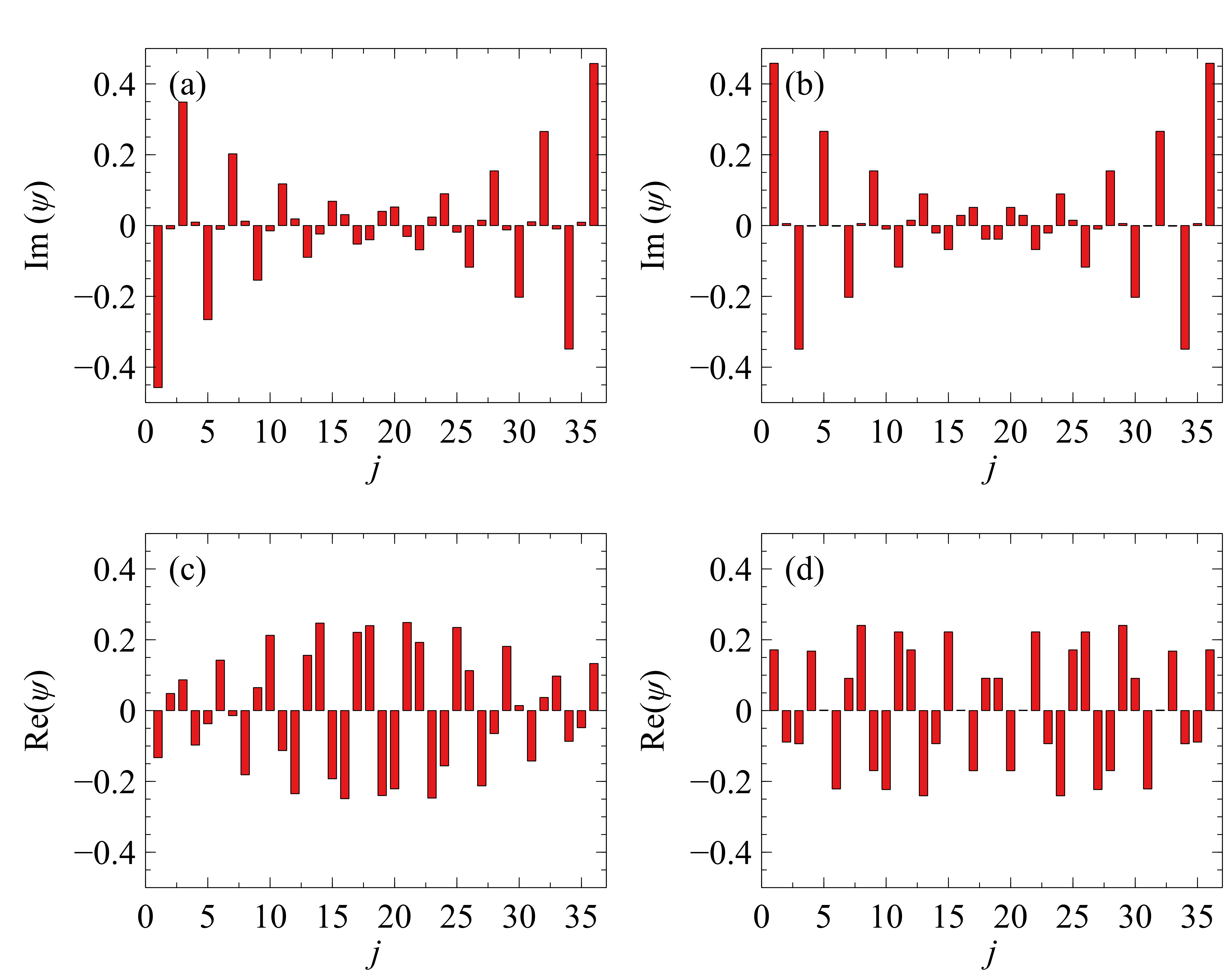}
    \caption{The normalized wavefunctions for the: (a) HOMO, (b) LUMO, (c) LUMO+1, and (d) LUMO+2 states of a carbyne chain containing 36 atoms, defined by the tight-binding parameters: $\alpha=-4.657$~eV, $\beta=-3.548$~eV and $\Delta_1=\Delta_2=-0.1$~eV. }
    \label{figure_wave}
\end{figure}

The transition dipole moment can be written in terms of the velocity matrix element using the relationship  
The velocity matrix element can be written in terms of the  transition dipole moment, using the relationship: 
$\left|\left\langle \psi_{f}\left|\boldsymbol{\mathrm{r}}\right|\psi_{i}\right\rangle \right|=\left|\left\langle \psi_{f}\left|\hbar\hat{\boldsymbol{\mathrm{v}}}\right|\psi_{i}\right\rangle /\left(E_{f}-E_{i}\right)\right|$.
For the case of $\Delta_1=\Delta_2=0$, this matrix element for the HOMO-LUMO transition can be expressed exactly using the approximate functions given in Eq.~(\ref{eq:exact_edge_wave}):
\begin{equation}
\left|\left\langle \psi_{f}\left|\boldsymbol{\mathrm{r}}\right|\psi_{i}\right\rangle \right|=\frac{c}{4}\left[N\frac{1+\left(\frac{\beta}{\alpha}\right)^{N}}{1-\left(\frac{\beta}{\alpha}\right)^{N}}-\frac{1+3\left(\frac{\beta}{\alpha}\right)^{2}}{1-\left(\frac{\beta}{\alpha}\right)^{2}}\right],
\label{eq:exact_RME}
\end{equation}
which linearly depends on $N$ in the long-chain limit. For transitions between an edge state and non-edge state, the transition dipole moment has a non-trivial dependence on $N$ (see Fig.~\ref{figure_v_mat}~(b)) since one function is exponential-like, the other plane-wave-like (see Fig.~\ref{figure_wave}). 

Let us now consider the case where both ends of the chain are terminated by gold, i.e., $\Delta_1=\Delta_2$. In this instance, our molecules possess mirror symmetry and the eigenfunctions are either even or odd, and within our eigenvalue numeration scheme, successive eigenvectors are of opposite parity. For an optical transition to be allowed the parity of the initial and final state must be different.  Since the eigenfunctions of the HOMO and LUMO levels are of differing parity (see Fig.~\ref{figure_wave}), optical transitions between them are allowed, and for long enough chains correspond to energies in the highly desirable THz range.  It should also be noted that if there is asymmetry between the on-site energies of the end atoms, the forbidden transitions for $\Delta_1=\Delta_2$ become allowed. 

In Fig.~\ref{figure_v_mat}~(a) we plot 
${\left|\left\langle \psi_{f}\left|\hbar\hat{\boldsymbol{\mathrm{v}}}\right|\psi_{i}\right\rangle \right|}^{2}/\left|
   \alpha c^{2}\left(E_{f}-E_{i}\right)\right|$ corresponding to the absolute value of the oscillator strength, normalized by a dimensionless constant $3\hbar^{2}/\left(2m_{e}c^{2}\alpha\right)$ (which is of the order of unity~\cite{harrison2012electronic}) for the transitions between levels close to the HOMO and LUMO, for $N=36$ and the tight-binding parameters $\alpha=-4.657$~eV, $\beta=-3.548$~eV~\cite{al2014electronic}, and $\Delta_1=\Delta_2=-0.1$~eV. This plot is done in the matrix style adopted from Ref.~\cite{buchs2021metallic}, which study another carbon-based system exhibiting THz transitions.
In panel~(b) of the same figure we plot the transition dipole moment for the same tight-binding parameters for a range of $N$. Here we have assumed that the metallic clusters terminating the edges results in the edge atom on-site energy differing from a carbon atom's on-site energy by 0.1 eV~\cite{valadbeigi2016ionization,kaiser2010experimental}. As can be seen from the figure, the strength of the LUMO-HOMO transition, i.e., between edge states, is of a comparable order of magnitude to the transitions associated with another mode.

\section*{THz Generation Scheme}
\begin{figure}[h]
    \centering
    \includegraphics[width=0.95\linewidth]{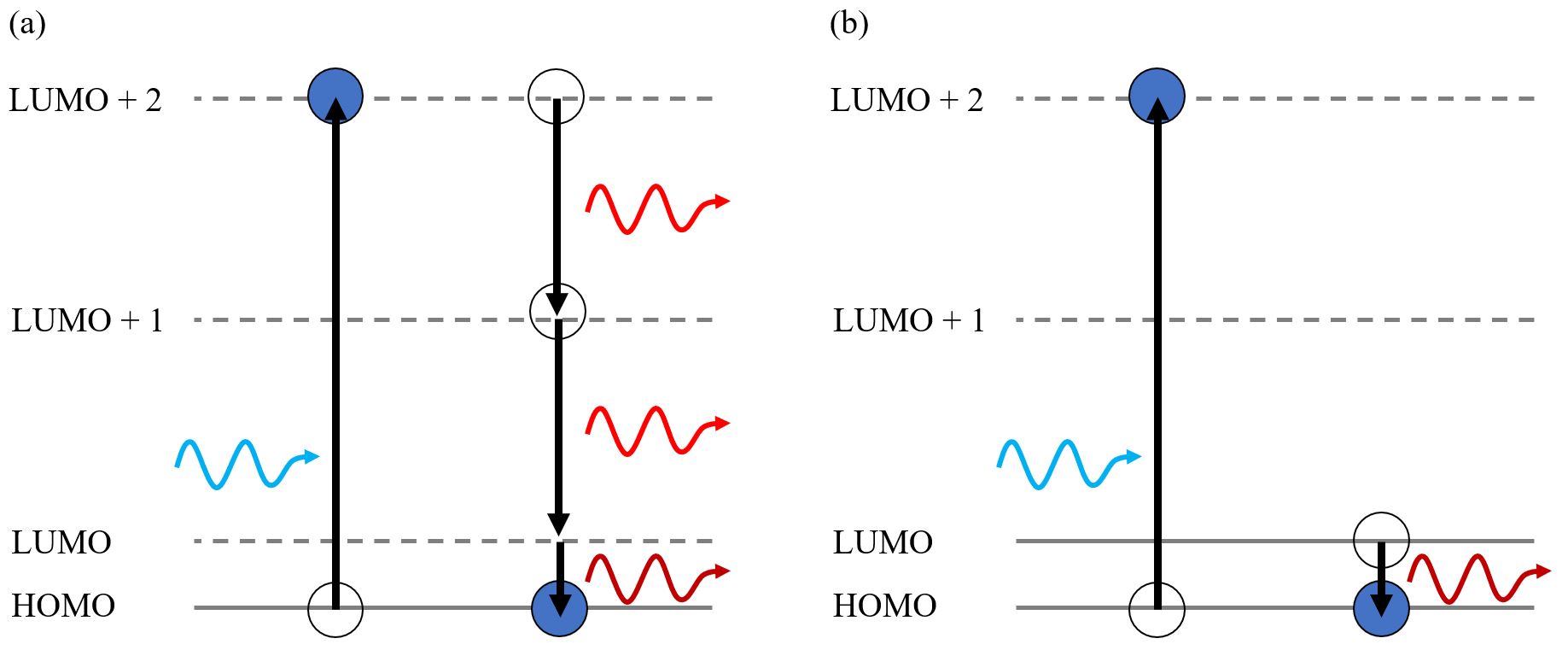}
    \caption{
    A schematic illustration of a THz generation scheme in (a) an undoped and (b) an n-doped polyyne carbyne chain. Here the occupied and unoccupied energy levels (prior to the photoexcitation) are depicted by the solid and dashed grey lines, respectively. In both cases a high-frequency optical excitation promotes an electron from the HOMO to the LUMO+2 energy level. Panel (a) shows the possible relaxation path to the ground state which would result in the emission of two low-frequency optical photons and one THz photon associated with the LUMO-HOMO transition. For the case of an n-doped carbyne chain, the unoccupied state created by the promotion of an electron out of the HOMO into the LUMO+2, can be filled by the relaxation of a LUMO electron via the emission of a THz photon. For a chain comprised of $N=36$ atoms, with tight-binding parameters: $\alpha=-4.657$~eV, $\beta=-3.548$~eV and $\Delta_1=\Delta_2=-0.1$~eV, 
    the excitation wavelength would be around $650$~nm while the emitted wavelengths from the LUMO+2 $\rightarrow$ LUMO+1, LUMO+1 $\rightarrow$ LUMO and LUMO $\rightarrow$ HOMO transitions are around 2250~nm, 930~nm and 42320~nm (7.08~THz) respectively.
}
    \label{figure_thz}
\end{figure}

In Fig.~\ref{figure_thz}~(a) we present one possible absorption and relaxation pathway which would result in the emission of THz radiation. First, a high-frequency optical excitation can be used to promote an electron from the HOMO to the LUMO+2 energy level (see Fig.~\ref{figure_thz}~(a)). Then the photoexcited carrier relaxes from the LUMO+2 to the LUMO+1 energy level, and then from the LUMO+1 to the LUMO energy level, each via the emission of a lower-frequency optical photon. Finally, the electron relaxes from the LUMO to the HOMO energy level via the emission of a THz photon. It should be noted that both the transition from the LUMO+2 to the LUMO energy level, and LUMO+1 to HOMO energy level are forbidden by symmetry (i.e., final and initial states are of the same parity). Since $\Delta_1=\Delta_2$, the energy difference between the HOMO$-2$ and LUMO levels is identical to the difference between the HOMO and LUMO+2 levels. Therefore, the same excitation used in Fig.~\ref{figure_thz} could, with equal probability, promote an electron from the HOMO$-2$ energy level into the LUMO energy level, allowing an electron in the HOMO$-1$ level to relax into the HOMO$-2$, which in turns allows an electron to relax from the HOMO level down to the HOMO$-1$: Thus allowing the photoexcited carrier in the LUMO level to relax into the HOMO level via the emission of a THz photon. For the case of n-doped polyyne chains, which is experimentally the more likely scenario, since gold complexes act as sources of free electrons, THz transitions can occur without the need of a multiphoton cascade process. For example, in Fig.~\ref{figure_thz}~(b) the polyyne chain is doped such that LUMO level is occupied. In this instance the promotion of an electron out of the HOMO into the LUMO+2 will leave an unoccupied state behind (thus creating a population inversion), which an electron from the LUMO can relax into via the emission of a THz photon.

Transitions in the THz frequency range are not uncommon in carbon compounds ~\cite{davies2008terahertz}. Still, the transition we propose in this work has important specific features that makes it promising for the realization of carbon-based THz lasers. Namely, Fig.~\ref{figure_v_mat} clearly demonstrates that the rate of the LUMO-HOMO transition is comparable to the rates of optical transitions in the system. This is of crucial importance for the realization of the inversion of electronic population needed to achieve lasing. Moreover, the frequency of this transition is tuneable in a wide range as it is strongly dependent on the length of the carbon chain. The optical pumping schemes illustrated in~Fig.~\ref{figure_thz}~(a,b) may allow for the realization of carbon-based THz lasers.

It should be noted that finite molecules possess several vibrational levels that can be coupled into the transition between electronic states. In our simple tight-binding model this would correspond to the tight-binding parameters being able to take a range of possible values, centered about the tight-binding parameter corresponding to the equilibrium bond position. Therefore, emission would occur across a broad range of wavelengths. It should also be noted that energy may be lost in internal conversion and vibrational relaxations, resulting in a lower frequency of fluorescent photons than our simple model predicts. The effects of Peierls distortion, bending, stretching and twisting of the chains can also be incorporated into an effective set of tight-binding parameters, and the nuanced effects due to each of the aforementioned stresses shall be a subject of future study. 

Although the employed simple model possesses an apparent symmetry of energy level positions with respect to the middle of the HOMO-LUMO gap (see Fig.~\ref{fig:chain_schematic}~(b)), it should be emphasized that our main conclusions do not rely on this symmetry. Asymmetry can be introduced into our model by taking into account the influence of the non-orthogonality of orbitals on adjacent atomic sites (see Appendix A). Much like in graphene in the vicinity of the Dirac points, in the proximity of the HOMO and LUMO levels of a finite chain, the modification to both the energy spectrum and the wave functions due to the overlap matrix can be treated as a small perturbation, which grows with increasing energy.

It should be emphasized that our results can be applied to other straight dimer chains, e.g., polyacetylene, via a suitable choice of hopping parameters, bond lengths, and onsite energies of the edge atoms (including their sign). The changes in parameters provide different numerical values of the band-gap and velocity matrix element; however, they do not change the results qualitatively.



\section*{Conclusion}
We have shown that carbyne offers a potential alternative to other low-dimensional forms of carbon for THz applications~\cite{Hartmann_2014_Rev,hartmann2019interband}, and predict that chains composed of 30 atoms or longer are promising candidates for the basis of THz emitters and detectors.
Although we have analyzed finite carbyne chains, the main results of our work can be generalized to various systems which exhibit topologically protected states separated by a finite distance. Indeed, much like in double quantum wells~\cite{hu1991feasibility,hartmann2020guided,hartmann2020bipolar}, these overlapping states may form symmetric and anti-symmetric functions, which support dipole optical transitions between them. Therefore providing the topologically protected states have a certain overlap, the system may support a THz energy gap between optically active states. Thus our proposed THz scheme could potentially be applied to topological materials with optically active bulk states, and geometrically constructed THz energy gaps associated with surface states.
We note also that conclusions of this theoretical work are qualitatively consistent with the recent experimental observations of low-temperature photoluminescence (PL) spectra of gold-capped polyyne chains containing from 10 to 24 carbon atoms ~\cite{kutrovskaya2020excitonic}.  We cautiously believe that carbon chains studied in ~\cite{kutrovskaya2020excitonic} were mostly n-doped, which is why the lowest energy PL transitions might have been between LUMO+1 and LUMO or even LUMO+2 and LUMO+1 energy levels. Indeed, while HOMO-LUMO transitions in these chains are in the infrared range, according to our calculations, the energies of optical transitions between various LUMO states as well as their dependences on the length of the chain appear to be in good qualitative agreement with the experimental data.






\section*{Acknowledgements}
This work was supported by the EU H2020 RISE project TERASSE (H2020-823878). RRH acknowledges financial support from URCO (14 F 1TAY20-1TAY21). The work of MEP was supported by the Russian Science Foundation (Project No. 20-12-00224), and he is also grateful for the hospitality of Westlake University, where a part of this work was done. The work of SK and AVK is supported by the Westlake University project No. 041020100118 and the Program 2018R01002 funded by Leading Innovative and Entrepreneur Team Introduction Program of Zhejiang. AK acknowledges the support from the Ministry of Science and Higher Education of the Russian Federation within the state assignment for Vladimir State University, project No. 0635-2020-0013.

\appendix

\section{Accounting for the wavefunction overlap}
The tight-binding Bloch function of an infinite polyyne chain can be written as
\begin{equation}
\Phi_{J}=\frac{1}{\sqrt{N}}\sum_{\boldsymbol{R}}^{N}e^{i\boldsymbol{k}\cdot\boldsymbol{R}}\varphi_{J}\left(\boldsymbol{r}-\boldsymbol{R}\right),\quad\left(J=1,2\right),
\end{equation}
where $\boldsymbol{R}$ denotes the atomic position and $\varphi_{J}$ the atomic wavefunction in state $J$. When the non-orthogonality
of orbitals on adjacent atomic sites is taken into account, the eigenvalue problem becomes $\hat{H}\Psi=E\hat{S}\Psi$~\cite{dresselhaus1998physical}, where $\hat{H}$ is the transfer integral matrix given in Eq.~(\ref{eq:Ham_con}) of the main text, and $\hat{S}$ is the overlap matrix, whose elements are given by $S_{JJ'}=\left\langle \Phi_{J}\mid\Phi_{J'}\right\rangle$. The secular equation, det($\hat{H}-E\hat{S}$)~=~0, thus becomes:
\begin{equation}
\left|\begin{array}{cc}
-E & \widetilde{f}^{\star}\\
\widetilde{f} & -E
\end{array}\right|=0,
\label{eq:asym_secular}
\end{equation}
where $\widetilde{f}=\left(\alpha-Es_{\alpha}\right)\exp\left(ikc_{s}\right)+\left(\beta-Es_{\beta}\right)\exp\left(-ikc_{l}\right)$. The asymmetry parameters are defined as $s_{\alpha}=\left\langle \varphi_{J'}\left(\boldsymbol{r}-\boldsymbol{R}-\boldsymbol{c}_{s}\right)\mid\varphi_{J}\left(\boldsymbol{r}-\boldsymbol{R}\right)\right\rangle$ and $s_{\beta}=\left\langle \varphi_{J'}\left(\boldsymbol{r}-\boldsymbol{R}-\boldsymbol{c}_{l}\right)\mid\varphi_{J}\left(\boldsymbol{r}-\boldsymbol{R}\right)\right\rangle$. Having non-zero values of $s_{\alpha}$ and $s_{\beta}$ breaks the symmetry between the positive and negative energy levels. Defining $qc=kc-\pi$, the solution to the secular equation near the band edge can be approximated by the expression:
\begin{equation}
E=\pm\sqrt{\frac{\left(\alpha-\beta\right)^{2}}{\left(s_{a}-s_{b}\mp1\right)^{2}}\mp\frac{\left(\beta\pm\delta\right)\left(\alpha\pm\delta\right)\left(qc\right)^{2}}{\left(s_{a}-s_{b}\mp1\right)^{3}}},
\end{equation}
where $\delta=\alpha s_{b}-\beta s_{a}$. Thus, much like for graphene near the apex of the Dirac cone~\cite{dresselhaus1998physical}, the modification to the energy spectrum due to the overlap matrix can be treated as a small perturbation. Therefore, the key results of the paper remain fundamentally unchanged in the presence of broken energy symmetry.

Finally, the secular equation of a finite chain which accounts for the non-orthogonality of orbitals on adjacent atomic sites, can be obtained by substituting $\alpha\rightarrow\left(\alpha-Es_{\alpha}\right)$ and $\beta\rightarrow\left(\beta-Es_{\beta}\right)$ into det($\mathbf{H}-E\mathbf{I}$)=0, where $\mathbf{H}$ is the Hamiltonian, Eq.~(\ref{eq:Ham_finite}) of the main text. This can be directly seen by comparing Eq.~(\ref{eq:asym_secular}) to the case where $\hat{S}=\textbf{I}$. However, it should be noted that when $E=0$, then $\widetilde{f}=f$; therefore the secular equation of the finite chain becomes identical to the case where the orbitals on adjacent atomic sites are orthogonal. Thus, for a symmetric chain in the infinite length limit, there is always a double-degenerate mid-gap state irrespective of the degree of non-orthogonality of the orbitals on adjacent atomic sites.

\bibliography{sample}

\begin{thebibliography}{46}%
\makeatletter
\providecommand \@ifxundefined [1]{%
 \@ifx{#1\undefined}
}%
\providecommand \@ifnum [1]{%
 \ifnum #1\expandafter \@firstoftwo
 \else \expandafter \@secondoftwo
 \fi
}%
\providecommand \@ifx [1]{%
 \ifx #1\expandafter \@firstoftwo
 \else \expandafter \@secondoftwo
 \fi
}%
\providecommand \natexlab [1]{#1}%
\providecommand \enquote  [1]{``#1''}%
\providecommand \bibnamefont  [1]{#1}%
\providecommand \bibfnamefont [1]{#1}%
\providecommand \citenamefont [1]{#1}%
\providecommand \href@noop [0]{\@secondoftwo}%
\providecommand \href [0]{\begingroup \@sanitize@url \@href}%
\providecommand \@href[1]{\@@startlink{#1}\@@href}%
\providecommand \@@href[1]{\endgroup#1\@@endlink}%
\providecommand \@sanitize@url [0]{\catcode `\\12\catcode `\$12\catcode
  `\&12\catcode `\#12\catcode `\^12\catcode `\_12\catcode `\%12\relax}%
\providecommand \@@startlink[1]{}%
\providecommand \@@endlink[0]{}%
\providecommand \url  [0]{\begingroup\@sanitize@url \@url }%
\providecommand \@url [1]{\endgroup\@href {#1}{\urlprefix }}%
\providecommand \urlprefix  [0]{URL }%
\providecommand \Eprint [0]{\href }%
\providecommand \doibase [0]{https://doi.org/}%
\providecommand \selectlanguage [0]{\@gobble}%
\providecommand \bibinfo  [0]{\@secondoftwo}%
\providecommand \bibfield  [0]{\@secondoftwo}%
\providecommand \translation [1]{[#1]}%
\providecommand \BibitemOpen [0]{}%
\providecommand \bibitemStop [0]{}%
\providecommand \bibitemNoStop [0]{.\EOS\space}%
\providecommand \EOS [0]{\spacefactor3000\relax}%
\providecommand \BibitemShut  [1]{\csname bibitem#1\endcsname}%
\let\auto@bib@innerbib\@empty
\bibitem [{\citenamefont {Kasatochin}\ \emph {et~al.}(1967)\citenamefont
  {Kasatochin}, \citenamefont {Sladov}, \citenamefont {Kudryavtseu},
  \citenamefont {Popov},\ and\ \citenamefont
  {Korshak}}]{kasatoch1967crystalline}%
  \BibitemOpen
  \bibfield  {author} {\bibinfo {author} {\bibfnamefont {V.~I.}\ \bibnamefont
  {Kasatochin}}, \bibinfo {author} {\bibfnamefont {A.~M.}\ \bibnamefont
  {Sladov}}, \bibinfo {author} {\bibfnamefont {Y.~P.}\ \bibnamefont
  {Kudryavtseu}}, \bibinfo {author} {\bibfnamefont {N.~M.}\ \bibnamefont
  {Popov}},\ and\ \bibinfo {author} {\bibfnamefont {V.~V.}\ \bibnamefont
  {Korshak}},\ }\bibfield  {title} {\bibinfo {title} {Crystalline forms of
  linear modification of carbon},\ }\href@noop {} {\bibfield  {journal}
  {\bibinfo  {journal} {Dokl. Akad. Nauk SSSR}\ }\textbf {\bibinfo {volume}
  {177}},\ \bibinfo {pages} {358} (\bibinfo {year} {1967})}\BibitemShut
  {NoStop}%
\bibitem [{\citenamefont {El~Goresy}\ and\ \citenamefont
  {Donnay}(1968)}]{el1968new}%
  \BibitemOpen
  \bibfield  {author} {\bibinfo {author} {\bibfnamefont {A.}~\bibnamefont
  {El~Goresy}}\ and\ \bibinfo {author} {\bibfnamefont {G.}~\bibnamefont
  {Donnay}},\ }\bibfield  {title} {\bibinfo {title} {A new allotropic form of
  carbon from the {R}ies crater},\ }\href@noop {} {\bibfield  {journal}
  {\bibinfo  {journal} {Science}\ }\textbf {\bibinfo {volume} {161}},\ \bibinfo
  {pages} {363} (\bibinfo {year} {1968})}\BibitemShut {NoStop}%
\bibitem [{\citenamefont {Whittaker}\ and\ \citenamefont
  {Kintner}(1969)}]{whittaker1969carbon}%
  \BibitemOpen
  \bibfield  {author} {\bibinfo {author} {\bibfnamefont {A.~G.}\ \bibnamefont
  {Whittaker}}\ and\ \bibinfo {author} {\bibfnamefont {P.~L.}\ \bibnamefont
  {Kintner}},\ }\bibfield  {title} {\bibinfo {title} {Carbon: Observations on
  the new allotropic form},\ }\href@noop {} {\bibfield  {journal} {\bibinfo
  {journal} {Science}\ }\textbf {\bibinfo {volume} {165}},\ \bibinfo {pages}
  {589} (\bibinfo {year} {1969})}\BibitemShut {NoStop}%
\bibitem [{\citenamefont {Sladkov}\ and\ \citenamefont
  {Kudryavtsev}(1969)}]{sladkov1969diamond}%
  \BibitemOpen
  \bibfield  {author} {\bibinfo {author} {\bibfnamefont {A.~M.}\ \bibnamefont
  {Sladkov}}\ and\ \bibinfo {author} {\bibfnamefont {Y.~P.}\ \bibnamefont
  {Kudryavtsev}},\ }\href@noop {} {\emph {\bibinfo {title} {Diamond, Graphite,
  Carbyne--The Allotropic Forms of Carbon (Almax, Grafit, Karbin--Allotropnye
  Formy Ugleroda)}}},\ \bibinfo {type} {Tech. Rep.}\ (\bibinfo  {institution}
  {Rand Corp Santa Monica Calif},\ \bibinfo {year} {1969})\BibitemShut
  {NoStop}%
\bibitem [{\citenamefont {Smith}\ and\ \citenamefont
  {Buseck}(1981)}]{smith1981graphitic}%
  \BibitemOpen
  \bibfield  {author} {\bibinfo {author} {\bibfnamefont {P.~P.}\ \bibnamefont
  {Smith}}\ and\ \bibinfo {author} {\bibfnamefont {P.~R.}\ \bibnamefont
  {Buseck}},\ }\bibfield  {title} {\bibinfo {title} {Graphitic carbon in the
  allende meteorite: a microstructural study},\ }\href@noop {} {\bibfield
  {journal} {\bibinfo  {journal} {Science}\ }\textbf {\bibinfo {volume}
  {212}},\ \bibinfo {pages} {322} (\bibinfo {year} {1981})}\BibitemShut
  {NoStop}%
\bibitem [{\citenamefont {Smith}\ and\ \citenamefont
  {Buseck}(1982)}]{smith1982carbyne}%
  \BibitemOpen
  \bibfield  {author} {\bibinfo {author} {\bibfnamefont {P.~P.~K.}\
  \bibnamefont {Smith}}\ and\ \bibinfo {author} {\bibfnamefont {P.~R.}\
  \bibnamefont {Buseck}},\ }\bibfield  {title} {\bibinfo {title} {Carbyne forms
  of carbon: do they exist?},\ }\href@noop {} {\bibfield  {journal} {\bibinfo
  {journal} {Science}\ }\textbf {\bibinfo {volume} {216}},\ \bibinfo {pages}
  {984} (\bibinfo {year} {1982})}\BibitemShut {NoStop}%
\bibitem [{\citenamefont {Heimann}\ \emph {et~al.}(1999)\citenamefont
  {Heimann}, \citenamefont {Evsyukov},\ and\ \citenamefont
  {Kavan}}]{heimann1999carbyne}%
  \BibitemOpen
  \bibfield  {author} {\bibinfo {author} {\bibfnamefont {R.~B.}\ \bibnamefont
  {Heimann}}, \bibinfo {author} {\bibfnamefont {S.~E.}\ \bibnamefont
  {Evsyukov}},\ and\ \bibinfo {author} {\bibfnamefont {L.}~\bibnamefont
  {Kavan}},\ }\href@noop {} {\emph {\bibinfo {title} {Carbyne and carbynoid
  structures}}},\ Vol.~\bibinfo {volume} {21}\ (\bibinfo  {publisher} {Springer
  Science \& Business Media},\ \bibinfo {year} {1999})\BibitemShut {NoStop}%
\bibitem [{\citenamefont {Landau}\ and\ \citenamefont
  {Lifshitz}(1980)}]{lifshitz2013statistical}%
  \BibitemOpen
  \bibfield  {author} {\bibinfo {author} {\bibfnamefont {L.}~\bibnamefont
  {Landau}}\ and\ \bibinfo {author} {\bibfnamefont {E.}~\bibnamefont
  {Lifshitz}},\ }\bibfield  {title} {\bibinfo {title} {Statistical physics,
  part 1: Volume 5 (course of theoretical physics, volume 5)},\ }\href@noop {}
  {\bibfield  {journal} {\bibinfo  {journal} {Publisher:
  Butterworth-Heinemann}\ }\textbf {\bibinfo {volume} {3}} (\bibinfo {year}
  {1980})}\BibitemShut {NoStop}%
\bibitem [{\citenamefont {Johnson}\ and\ \citenamefont
  {Walton}(1972)}]{johnson1972silylation}%
  \BibitemOpen
  \bibfield  {author} {\bibinfo {author} {\bibfnamefont {T.~R.}\ \bibnamefont
  {Johnson}}\ and\ \bibinfo {author} {\bibfnamefont {D.~R.~M.}\ \bibnamefont
  {Walton}},\ }\bibfield  {title} {\bibinfo {title} {Silylation as a protective
  method in acetylene chemistry: Polyyne chain extensions using the reagents,
  {E}t$_{3}${S}i({C}$\equiv${C})$_{m}${H} (m=1, 2, 4) in mixed oxidative
  couplings},\ }\href@noop {} {\bibfield  {journal} {\bibinfo  {journal}
  {Tetrahedron}\ }\textbf {\bibinfo {volume} {28}},\ \bibinfo {pages} {5221}
  (\bibinfo {year} {1972})}\BibitemShut {NoStop}%
\bibitem [{\citenamefont {Gibtner}\ \emph {et~al.}(2002)\citenamefont
  {Gibtner}, \citenamefont {Hampel}, \citenamefont {Gisselbrecht},\ and\
  \citenamefont {Hirsch}}]{gibtner2002end}%
  \BibitemOpen
  \bibfield  {author} {\bibinfo {author} {\bibfnamefont {T.}~\bibnamefont
  {Gibtner}}, \bibinfo {author} {\bibfnamefont {F.}~\bibnamefont {Hampel}},
  \bibinfo {author} {\bibfnamefont {J.-P.}\ \bibnamefont {Gisselbrecht}},\ and\
  \bibinfo {author} {\bibfnamefont {A.}~\bibnamefont {Hirsch}},\ }\bibfield
  {title} {\bibinfo {title} {End-cap stabilized oligoynes: model compounds for
  the linear sp carbon allotrope carbyne},\ }\href@noop {} {\bibfield
  {journal} {\bibinfo  {journal} {Chem. Eur. J}\ }\textbf {\bibinfo {volume}
  {8}},\ \bibinfo {pages} {408} (\bibinfo {year} {2002})}\BibitemShut {NoStop}%
\bibitem [{\citenamefont {Chalifoux}\ and\ \citenamefont
  {Tykwinski}(2010)}]{chalifoux2010synthesis}%
  \BibitemOpen
  \bibfield  {author} {\bibinfo {author} {\bibfnamefont {W.~A.}\ \bibnamefont
  {Chalifoux}}\ and\ \bibinfo {author} {\bibfnamefont {R.~R.}\ \bibnamefont
  {Tykwinski}},\ }\bibfield  {title} {\bibinfo {title} {Synthesis of polyynes
  to model the sp-carbon allotrope carbyne},\ }\href@noop {} {\bibfield
  {journal} {\bibinfo  {journal} {Nature chemistry}\ }\textbf {\bibinfo
  {volume} {2}},\ \bibinfo {pages} {967} (\bibinfo {year} {2010})}\BibitemShut
  {NoStop}%
\bibitem [{\citenamefont {Nishide}\ \emph {et~al.}(2006)\citenamefont
  {Nishide}, \citenamefont {Dohi}, \citenamefont {Wakabayashi}, \citenamefont
  {Nishibori}, \citenamefont {Aoyagi}, \citenamefont {Ishida}, \citenamefont
  {Kikuchi}, \citenamefont {Kitaura}, \citenamefont {Sugai}, \citenamefont
  {Sakata} \emph {et~al.}}]{nishide2006single}%
  \BibitemOpen
  \bibfield  {author} {\bibinfo {author} {\bibfnamefont {D.}~\bibnamefont
  {Nishide}}, \bibinfo {author} {\bibfnamefont {H.}~\bibnamefont {Dohi}},
  \bibinfo {author} {\bibfnamefont {T.}~\bibnamefont {Wakabayashi}}, \bibinfo
  {author} {\bibfnamefont {E.}~\bibnamefont {Nishibori}}, \bibinfo {author}
  {\bibfnamefont {S.}~\bibnamefont {Aoyagi}}, \bibinfo {author} {\bibfnamefont
  {M.}~\bibnamefont {Ishida}}, \bibinfo {author} {\bibfnamefont
  {S.}~\bibnamefont {Kikuchi}}, \bibinfo {author} {\bibfnamefont
  {R.}~\bibnamefont {Kitaura}}, \bibinfo {author} {\bibfnamefont
  {T.}~\bibnamefont {Sugai}}, \bibinfo {author} {\bibfnamefont
  {M.}~\bibnamefont {Sakata}}, \emph {et~al.},\ }\bibfield  {title} {\bibinfo
  {title} {Single-wall carbon nanotubes encaging linear chain {C}10{H}2 polyyne
  molecules inside},\ }\href@noop {} {\bibfield  {journal} {\bibinfo  {journal}
  {Chem. Phys. Lett}\ }\textbf {\bibinfo {volume} {428}},\ \bibinfo {pages}
  {356} (\bibinfo {year} {2006})}\BibitemShut {NoStop}%
\bibitem [{\citenamefont {Shi}\ \emph {et~al.}(2016)\citenamefont {Shi},
  \citenamefont {Rohringer}, \citenamefont {Suenaga}, \citenamefont {Niimi},
  \citenamefont {Kotakoski}, \citenamefont {Meyer}, \citenamefont {Peterlik},
  \citenamefont {Wanko}, \citenamefont {Cahangirov}, \citenamefont {Rubio}
  \emph {et~al.}}]{shi2016confined}%
  \BibitemOpen
  \bibfield  {author} {\bibinfo {author} {\bibfnamefont {L.}~\bibnamefont
  {Shi}}, \bibinfo {author} {\bibfnamefont {P.}~\bibnamefont {Rohringer}},
  \bibinfo {author} {\bibfnamefont {K.}~\bibnamefont {Suenaga}}, \bibinfo
  {author} {\bibfnamefont {Y.}~\bibnamefont {Niimi}}, \bibinfo {author}
  {\bibfnamefont {J.}~\bibnamefont {Kotakoski}}, \bibinfo {author}
  {\bibfnamefont {J.~C.}\ \bibnamefont {Meyer}}, \bibinfo {author}
  {\bibfnamefont {H.}~\bibnamefont {Peterlik}}, \bibinfo {author}
  {\bibfnamefont {M.}~\bibnamefont {Wanko}}, \bibinfo {author} {\bibfnamefont
  {S.}~\bibnamefont {Cahangirov}}, \bibinfo {author} {\bibfnamefont
  {A.}~\bibnamefont {Rubio}}, \emph {et~al.},\ }\bibfield  {title} {\bibinfo
  {title} {Confined linear carbon chains as a route to bulk carbyne},\
  }\href@noop {} {\bibfield  {journal} {\bibinfo  {journal} {Nature materials}\
  }\textbf {\bibinfo {volume} {15}},\ \bibinfo {pages} {634} (\bibinfo {year}
  {2016})}\BibitemShut {NoStop}%
\bibitem [{\citenamefont {Cataldo}(2004)}]{cataldo2004synthesis}%
  \BibitemOpen
  \bibfield  {author} {\bibinfo {author} {\bibfnamefont {F.}~\bibnamefont
  {Cataldo}},\ }\bibfield  {title} {\bibinfo {title} {Synthesis of polyynes in
  a submerged electric arc in organic solvents},\ }\href
  {https://doi.org/https://doi.org/10.1016/j.carbon.2003.10.016} {\bibfield
  {journal} {\bibinfo  {journal} {Carbon}\ }\textbf {\bibinfo {volume} {42}},\
  \bibinfo {pages} {129} (\bibinfo {year} {2004})}\BibitemShut {NoStop}%
\bibitem [{\citenamefont {Pan}\ \emph {et~al.}(2015)\citenamefont {Pan},
  \citenamefont {Xiao}, \citenamefont {Li}, \citenamefont {Liu}, \citenamefont
  {Wang},\ and\ \citenamefont {Yang}}]{pan2015carbyne}%
  \BibitemOpen
  \bibfield  {author} {\bibinfo {author} {\bibfnamefont {B.}~\bibnamefont
  {Pan}}, \bibinfo {author} {\bibfnamefont {J.}~\bibnamefont {Xiao}}, \bibinfo
  {author} {\bibfnamefont {J.}~\bibnamefont {Li}}, \bibinfo {author}
  {\bibfnamefont {P.}~\bibnamefont {Liu}}, \bibinfo {author} {\bibfnamefont
  {C.}~\bibnamefont {Wang}},\ and\ \bibinfo {author} {\bibfnamefont
  {G.}~\bibnamefont {Yang}},\ }\bibfield  {title} {\bibinfo {title} {Carbyne
  with finite length: The one-dimensional sp carbon},\ }\href@noop {}
  {\bibfield  {journal} {\bibinfo  {journal} {Sci. Adv}\ }\textbf {\bibinfo
  {volume} {1}} (\bibinfo {year} {2015})}\BibitemShut {NoStop}%
\bibitem [{\citenamefont {Kucherik}\ \emph {et~al.}(2016)\citenamefont
  {Kucherik}, \citenamefont {Arakelian}, \citenamefont {Garnov}, \citenamefont
  {Kutrovskaya}, \citenamefont {Nogtev}, \citenamefont {Osipov},\ and\
  \citenamefont {Khor'kov}}]{kucherik2016two}%
  \BibitemOpen
  \bibfield  {author} {\bibinfo {author} {\bibfnamefont {A.~O.}\ \bibnamefont
  {Kucherik}}, \bibinfo {author} {\bibfnamefont {S.~M.}\ \bibnamefont
  {Arakelian}}, \bibinfo {author} {\bibfnamefont {S.~V.}\ \bibnamefont
  {Garnov}}, \bibinfo {author} {\bibfnamefont {S.~V.}\ \bibnamefont
  {Kutrovskaya}}, \bibinfo {author} {\bibfnamefont {D.~S.}\ \bibnamefont
  {Nogtev}}, \bibinfo {author} {\bibfnamefont {A.~V.}\ \bibnamefont {Osipov}},\
  and\ \bibinfo {author} {\bibfnamefont {K.~S.}\ \bibnamefont {Khor'kov}},\
  }\bibfield  {title} {\bibinfo {title} {Two-stage laser-induced synthesis of
  linear carbon chains},\ }\href@noop {} {\bibfield  {journal} {\bibinfo
  {journal} {Quantum Electronics}\ }\textbf {\bibinfo {volume} {46}},\ \bibinfo
  {pages} {627} (\bibinfo {year} {2016})}\BibitemShut {NoStop}%
\bibitem [{\citenamefont {Kutrovskaya}\ \emph {et~al.}(2020)\citenamefont
  {Kutrovskaya}, \citenamefont {Osipov}, \citenamefont {Baryshev},
  \citenamefont {Zasedatelev}, \citenamefont {Samyshkin}, \citenamefont
  {Demirchyan}, \citenamefont {Pulci}, \citenamefont {Grassano}, \citenamefont
  {Gontrani}, \citenamefont {Hartmann}, \citenamefont {Portnoi}, \citenamefont
  {Kucherik}, \citenamefont {Lagoudakis},\ and\ \citenamefont
  {Kavokin}}]{kutrovskaya2020excitonic}%
  \BibitemOpen
  \bibfield  {author} {\bibinfo {author} {\bibfnamefont {S.}~\bibnamefont
  {Kutrovskaya}}, \bibinfo {author} {\bibfnamefont {A.}~\bibnamefont {Osipov}},
  \bibinfo {author} {\bibfnamefont {S.}~\bibnamefont {Baryshev}}, \bibinfo
  {author} {\bibfnamefont {A.}~\bibnamefont {Zasedatelev}}, \bibinfo {author}
  {\bibfnamefont {V.}~\bibnamefont {Samyshkin}}, \bibinfo {author}
  {\bibfnamefont {S.}~\bibnamefont {Demirchyan}}, \bibinfo {author}
  {\bibfnamefont {O.}~\bibnamefont {Pulci}}, \bibinfo {author} {\bibfnamefont
  {D.}~\bibnamefont {Grassano}}, \bibinfo {author} {\bibfnamefont
  {L.}~\bibnamefont {Gontrani}}, \bibinfo {author} {\bibfnamefont {R.~R.}\
  \bibnamefont {Hartmann}}, \bibinfo {author} {\bibfnamefont {M.~E.}\
  \bibnamefont {Portnoi}}, \bibinfo {author} {\bibfnamefont {A.}~\bibnamefont
  {Kucherik}}, \bibinfo {author} {\bibfnamefont {P.~G.}\ \bibnamefont
  {Lagoudakis}},\ and\ \bibinfo {author} {\bibfnamefont {A.}~\bibnamefont
  {Kavokin}},\ }\bibfield  {title} {\bibinfo {title} {Excitonic fine structure
  in emission of linear carbon chains},\ }\href@noop {} {\bibfield  {journal}
  {\bibinfo  {journal} {Nano Lett.}\ }\textbf {\bibinfo {volume} {20}},\
  \bibinfo {pages} {6502} (\bibinfo {year} {2020})}\BibitemShut {NoStop}%
\bibitem [{\citenamefont {Bianco}\ \emph {et~al.}(2018)\citenamefont {Bianco},
  \citenamefont {Chen}, \citenamefont {Chen}, \citenamefont {Ghoshal},
  \citenamefont {Hurt}, \citenamefont {Kim}, \citenamefont {Koratkar},
  \citenamefont {Meunier},\ and\ \citenamefont {Terrones}}]{bianco2018carbon}%
  \BibitemOpen
  \bibfield  {author} {\bibinfo {author} {\bibfnamefont {A.}~\bibnamefont
  {Bianco}}, \bibinfo {author} {\bibfnamefont {Y.}~\bibnamefont {Chen}},
  \bibinfo {author} {\bibfnamefont {Y.}~\bibnamefont {Chen}}, \bibinfo {author}
  {\bibfnamefont {D.}~\bibnamefont {Ghoshal}}, \bibinfo {author} {\bibfnamefont
  {R.~H.}\ \bibnamefont {Hurt}}, \bibinfo {author} {\bibfnamefont {Y.~A.}\
  \bibnamefont {Kim}}, \bibinfo {author} {\bibfnamefont {N.}~\bibnamefont
  {Koratkar}}, \bibinfo {author} {\bibfnamefont {V.}~\bibnamefont {Meunier}},\
  and\ \bibinfo {author} {\bibfnamefont {M.}~\bibnamefont {Terrones}},\
  }\bibfield  {title} {\bibinfo {title} {A carbon science perspective in 2018:
  Current achievements and future challenges},\ }\href@noop {} {\bibfield
  {journal} {\bibinfo  {journal} {Carbon}\ }\textbf {\bibinfo {volume} {132}},\
  \bibinfo {pages} {785} (\bibinfo {year} {2018})}\BibitemShut {NoStop}%
\bibitem [{\citenamefont {Hartmann}\ and\ \citenamefont
  {Portnoi}(2020{\natexlab{a}})}]{hartmann2020guided}%
  \BibitemOpen
  \bibfield  {author} {\bibinfo {author} {\bibfnamefont {R.~R.}\ \bibnamefont
  {Hartmann}}\ and\ \bibinfo {author} {\bibfnamefont {M.~E.}\ \bibnamefont
  {Portnoi}},\ }\bibfield  {title} {\bibinfo {title} {Guided modes and
  terahertz transitions for two-dimensional {D}irac fermions in a smooth
  double-well potential},\ }\href {https://doi.org/10.1103/PhysRevA.102.052229}
  {\bibfield  {journal} {\bibinfo  {journal} {Phys. Rev. A}\ }\textbf {\bibinfo
  {volume} {102}},\ \bibinfo {pages} {052229} (\bibinfo {year}
  {2020}{\natexlab{a}})}\BibitemShut {NoStop}%
\bibitem [{\citenamefont {Hartmann}\ and\ \citenamefont
  {Portnoi}(2020{\natexlab{b}})}]{hartmann2020bipolar}%
  \BibitemOpen
  \bibfield  {author} {\bibinfo {author} {\bibfnamefont {R.~R.}\ \bibnamefont
  {Hartmann}}\ and\ \bibinfo {author} {\bibfnamefont {M.~E.}\ \bibnamefont
  {Portnoi}},\ }\bibfield  {title} {\bibinfo {title} {Bipolar electron
  waveguides in graphene},\ }\href
  {https://doi.org/10.1103/PhysRevB.102.155421} {\bibfield  {journal} {\bibinfo
   {journal} {Phys. Rev. B}\ }\textbf {\bibinfo {volume} {102}},\ \bibinfo
  {pages} {155421} (\bibinfo {year} {2020}{\natexlab{b}})}\BibitemShut
  {NoStop}%
\bibitem [{\citenamefont {Al-Backri}\ \emph {et~al.}(2014)\citenamefont
  {Al-Backri}, \citenamefont {Z{\'o}lyomi},\ and\ \citenamefont
  {Lambert}}]{al2014electronic}%
  \BibitemOpen
  \bibfield  {author} {\bibinfo {author} {\bibfnamefont {A.}~\bibnamefont
  {Al-Backri}}, \bibinfo {author} {\bibfnamefont {V.}~\bibnamefont
  {Z{\'o}lyomi}},\ and\ \bibinfo {author} {\bibfnamefont {C.~J.}\ \bibnamefont
  {Lambert}},\ }\bibfield  {title} {\bibinfo {title} {Electronic properties of
  linear carbon chains: Resolving the controversy},\ }\href@noop {} {\bibfield
  {journal} {\bibinfo  {journal} {J. Chem. Phys}\ }\textbf {\bibinfo {volume}
  {140}},\ \bibinfo {pages} {104306} (\bibinfo {year} {2014})}\BibitemShut
  {NoStop}%
\bibitem [{\citenamefont {Rice}\ \emph {et~al.}(1986)\citenamefont {Rice},
  \citenamefont {Phillpot}, \citenamefont {Bishop},\ and\ \citenamefont
  {Campbell}}]{rice1986solitons}%
  \BibitemOpen
  \bibfield  {author} {\bibinfo {author} {\bibfnamefont {M.~J.}\ \bibnamefont
  {Rice}}, \bibinfo {author} {\bibfnamefont {S.~R.}\ \bibnamefont {Phillpot}},
  \bibinfo {author} {\bibfnamefont {A.~R.}\ \bibnamefont {Bishop}},\ and\
  \bibinfo {author} {\bibfnamefont {D.~K.}\ \bibnamefont {Campbell}},\
  }\bibfield  {title} {\bibinfo {title} {Solitons, polarons, and phonons in the
  infinite polyyne chain},\ }\href {https://doi.org/10.1103/PhysRevB.34.4139}
  {\bibfield  {journal} {\bibinfo  {journal} {Phys. Rev. B}\ }\textbf {\bibinfo
  {volume} {34}},\ \bibinfo {pages} {4139} (\bibinfo {year}
  {1986})}\BibitemShut {NoStop}%
\bibitem [{\citenamefont {Dresselhaus}\ \emph {et~al.}(1998)\citenamefont
  {Dresselhaus}, \citenamefont {Dresselhaus},\ and\ \citenamefont
  {Saito}}]{dresselhaus1998physical}%
  \BibitemOpen
  \bibfield  {author} {\bibinfo {author} {\bibfnamefont {G.}~\bibnamefont
  {Dresselhaus}}, \bibinfo {author} {\bibfnamefont {M.~S.}\ \bibnamefont
  {Dresselhaus}},\ and\ \bibinfo {author} {\bibfnamefont {R.}~\bibnamefont
  {Saito}},\ }\href@noop {} {\emph {\bibinfo {title} {Physical properties of
  carbon nanotubes}}}\ (\bibinfo  {publisher} {World Scientific},\ \bibinfo
  {year} {1998})\BibitemShut {NoStop}%
\bibitem [{\citenamefont {Hartmann}\ \emph {et~al.}(2014)\citenamefont
  {Hartmann}, \citenamefont {Kono},\ and\ \citenamefont
  {Portnoi}}]{Hartmann_2014_Rev}%
  \BibitemOpen
  \bibfield  {author} {\bibinfo {author} {\bibfnamefont {R.~R.}\ \bibnamefont
  {Hartmann}}, \bibinfo {author} {\bibfnamefont {J.}~\bibnamefont {Kono}},\
  and\ \bibinfo {author} {\bibfnamefont {M.~E.}\ \bibnamefont {Portnoi}},\
  }\bibfield  {title} {\bibinfo {title} {Terahertz science and technology of
  carbon nanomaterials},\ }\href
  {https://doi.org/10.1088/0957-4484/25/32/322001} {\bibfield  {journal}
  {\bibinfo  {journal} {Nanotechnology}\ }\textbf {\bibinfo {volume} {25}},\
  \bibinfo {pages} {322001} (\bibinfo {year} {2014})}\BibitemShut {NoStop}%
\bibitem [{\citenamefont {Castro~Neto}\ \emph {et~al.}(2009)\citenamefont
  {Castro~Neto}, \citenamefont {Guinea}, \citenamefont {Peres}, \citenamefont
  {Novoselov},\ and\ \citenamefont {Geim}}]{neto2009electronic}%
  \BibitemOpen
  \bibfield  {author} {\bibinfo {author} {\bibfnamefont {A.~H.}\ \bibnamefont
  {Castro~Neto}}, \bibinfo {author} {\bibfnamefont {F.}~\bibnamefont {Guinea}},
  \bibinfo {author} {\bibfnamefont {N.~M.~R.}\ \bibnamefont {Peres}}, \bibinfo
  {author} {\bibfnamefont {K.~S.}\ \bibnamefont {Novoselov}},\ and\ \bibinfo
  {author} {\bibfnamefont {A.~K.}\ \bibnamefont {Geim}},\ }\bibfield  {title}
  {\bibinfo {title} {The electronic properties of graphene},\ }\href
  {https://doi.org/10.1103/RevModPhys.81.109} {\bibfield  {journal} {\bibinfo
  {journal} {Rev. Mod. Phys.}\ }\textbf {\bibinfo {volume} {81}},\ \bibinfo
  {pages} {109} (\bibinfo {year} {2009})}\BibitemShut {NoStop}%
\bibitem [{\citenamefont {Losonczi}(1992)}]{losonczi1992eigenvalues}%
  \BibitemOpen
  \bibfield  {author} {\bibinfo {author} {\bibfnamefont {L.}~\bibnamefont
  {Losonczi}},\ }\bibfield  {title} {\bibinfo {title} {Eigenvalues and
  eigenvectors of some tridiagonal matrices},\ }\href@noop {} {\bibfield
  {journal} {\bibinfo  {journal} {Acta Math. Hungar.}\ }\textbf {\bibinfo
  {volume} {60}},\ \bibinfo {pages} {309} (\bibinfo {year} {1992})}\BibitemShut
  {NoStop}%
\bibitem [{\citenamefont {Yueh}(2005)}]{yueh2005eigenvalues}%
  \BibitemOpen
  \bibfield  {author} {\bibinfo {author} {\bibfnamefont {W.-C.}\ \bibnamefont
  {Yueh}},\ }\bibfield  {title} {\bibinfo {title} {Eigenvalues of several
  tridiagonal matrices},\ }\href@noop {} {\bibfield  {journal} {\bibinfo
  {journal} {Applied Mathematics E-Notes}\ }\textbf {\bibinfo {volume} {5}},\
  \bibinfo {pages} {66} (\bibinfo {year} {2005})}\BibitemShut {NoStop}%
\bibitem [{\citenamefont {Kouachi}(2006)}]{kouachi2006eigenvalues}%
  \BibitemOpen
  \bibfield  {author} {\bibinfo {author} {\bibfnamefont {S.}~\bibnamefont
  {Kouachi}},\ }\bibfield  {title} {\bibinfo {title} {Eigenvalues and
  eigenvectors of tridiagonal matrices},\ }\href@noop {} {\bibfield  {journal}
  {\bibinfo  {journal} {Electron. J. Linear Algebra}\ }\textbf {\bibinfo
  {volume} {15}},\ \bibinfo {pages} {115} (\bibinfo {year} {2006})}\BibitemShut
  {NoStop}%
\bibitem [{\citenamefont {da~Fonseca}(2007)}]{da2007characteristic}%
  \BibitemOpen
  \bibfield  {author} {\bibinfo {author} {\bibfnamefont {C.~M.}\ \bibnamefont
  {da~Fonseca}},\ }\bibfield  {title} {\bibinfo {title} {The characteristic
  polynomial of some perturbed tridiagonal k-toeplitz matrices},\ }\href@noop
  {} {\bibfield  {journal} {\bibinfo  {journal} {Appl. Math. Sci}\ }\textbf
  {\bibinfo {volume} {1}},\ \bibinfo {pages} {59} (\bibinfo {year}
  {2007})}\BibitemShut {NoStop}%
\bibitem [{\citenamefont {Willms}(2008)}]{willms2008analytic}%
  \BibitemOpen
  \bibfield  {author} {\bibinfo {author} {\bibfnamefont {A.~R.}\ \bibnamefont
  {Willms}},\ }\bibfield  {title} {\bibinfo {title} {Analytic results for the
  eigenvalues of certain tridiagonal matrices},\ }\href@noop {} {\bibfield
  {journal} {\bibinfo  {journal} {{SIAM} J. Matrix Anal. Appl}\ }\textbf
  {\bibinfo {volume} {30}},\ \bibinfo {pages} {639} (\bibinfo {year}
  {2008})}\BibitemShut {NoStop}%
\bibitem [{\citenamefont {Kouachi}(2008)}]{kouachi2008eigenvalues}%
  \BibitemOpen
  \bibfield  {author} {\bibinfo {author} {\bibfnamefont {S.}~\bibnamefont
  {Kouachi}},\ }\bibfield  {title} {\bibinfo {title} {Eigenvalues and
  eigenvectors of some tridiagonal matrices with non-constant diagonal
  entries},\ }\href@noop {} {\bibfield  {journal} {\bibinfo  {journal}
  {Applicationes mathematicae}\ }\textbf {\bibinfo {volume} {35}},\ \bibinfo
  {pages} {107} (\bibinfo {year} {2008})}\BibitemShut {NoStop}%
\bibitem [{\citenamefont {Da~Fonseca}\ and\ \citenamefont
  {Kowalenko}(2020)}]{da2019eigenpairs}%
  \BibitemOpen
  \bibfield  {author} {\bibinfo {author} {\bibfnamefont {C.~M.}\ \bibnamefont
  {Da~Fonseca}}\ and\ \bibinfo {author} {\bibfnamefont {V.}~\bibnamefont
  {Kowalenko}},\ }\bibfield  {title} {\bibinfo {title} {Eigenpairs of a family
  of tridiagonal matrices: three decades later},\ }\href@noop {} {\bibfield
  {journal} {\bibinfo  {journal} {Acta Math. Hungar.}\ }\textbf {\bibinfo
  {volume} {160}},\ \bibinfo {pages} {376} (\bibinfo {year}
  {2020})}\BibitemShut {NoStop}%
\bibitem [{\citenamefont {Deretzis}\ and\ \citenamefont
  {La~Magna}(2011)}]{deretzis2011coherent}%
  \BibitemOpen
  \bibfield  {author} {\bibinfo {author} {\bibfnamefont {I.}~\bibnamefont
  {Deretzis}}\ and\ \bibinfo {author} {\bibfnamefont {A.}~\bibnamefont
  {La~Magna}},\ }\bibfield  {title} {\bibinfo {title} {Coherent electron
  transport in quasi one-dimensional carbon-based systems},\ }\href@noop {}
  {\bibfield  {journal} {\bibinfo  {journal} {Eur. Phys. J. B}\ }\textbf
  {\bibinfo {volume} {81}},\ \bibinfo {pages} {15} (\bibinfo {year}
  {2011})}\BibitemShut {NoStop}%
\bibitem [{\citenamefont {Nakada}\ \emph {et~al.}(1996)\citenamefont {Nakada},
  \citenamefont {Fujita}, \citenamefont {Dresselhaus},\ and\ \citenamefont
  {Dresselhaus}}]{PhysRevB.54.17954}%
  \BibitemOpen
  \bibfield  {author} {\bibinfo {author} {\bibfnamefont {K.}~\bibnamefont
  {Nakada}}, \bibinfo {author} {\bibfnamefont {M.}~\bibnamefont {Fujita}},
  \bibinfo {author} {\bibfnamefont {G.}~\bibnamefont {Dresselhaus}},\ and\
  \bibinfo {author} {\bibfnamefont {M.~S.}\ \bibnamefont {Dresselhaus}},\
  }\bibfield  {title} {\bibinfo {title} {Edge state in graphene ribbons:
  {N}anometer size effect and edge shape dependence},\ }\href@noop {}
  {\bibfield  {journal} {\bibinfo  {journal} {Phys. Rev. B}\ }\textbf {\bibinfo
  {volume} {54}},\ \bibinfo {pages} {17954} (\bibinfo {year}
  {1996})}\BibitemShut {NoStop}%
\bibitem [{\citenamefont {St{e}\'{s}licka}\ \emph {et~al.}(2002)\citenamefont
  {St{e}\'{s}licka}, \citenamefont {Kucharczyk}, \citenamefont {Akjouj},
  \citenamefont {Djafari-Rouhani}, \citenamefont {Dobrzynski},\ and\
  \citenamefont {Davison}}]{STESLICKA200293}%
  \BibitemOpen
  \bibfield  {author} {\bibinfo {author} {\bibfnamefont {M.}~\bibnamefont
  {St{e}\'{s}licka}}, \bibinfo {author} {\bibfnamefont {R.}~\bibnamefont
  {Kucharczyk}}, \bibinfo {author} {\bibfnamefont {A.}~\bibnamefont {Akjouj}},
  \bibinfo {author} {\bibfnamefont {B.}~\bibnamefont {Djafari-Rouhani}},
  \bibinfo {author} {\bibfnamefont {L.}~\bibnamefont {Dobrzynski}},\ and\
  \bibinfo {author} {\bibfnamefont {S.~G.}\ \bibnamefont {Davison}},\
  }\bibfield  {title} {\bibinfo {title} {Localised electronic states in
  semiconductor superlattices},\ }\href
  {https://doi.org/https://doi.org/10.1016/S0167-5729(02)00052-3} {\bibfield
  {journal} {\bibinfo  {journal} {Surf. Sci. Rep}\ }\textbf {\bibinfo {volume}
  {47}},\ \bibinfo {pages} {93 } (\bibinfo {year} {2002})}\BibitemShut
  {NoStop}%
\bibitem [{\citenamefont {Su}\ \emph {et~al.}(1979)\citenamefont {Su},
  \citenamefont {Schrieffer},\ and\ \citenamefont
  {Heeger}}]{PhysRevLett.42.1698}%
  \BibitemOpen
  \bibfield  {author} {\bibinfo {author} {\bibfnamefont {W.~P.}\ \bibnamefont
  {Su}}, \bibinfo {author} {\bibfnamefont {J.~R.}\ \bibnamefont {Schrieffer}},\
  and\ \bibinfo {author} {\bibfnamefont {A.~J.}\ \bibnamefont {Heeger}},\
  }\bibfield  {title} {\bibinfo {title} {Solitons in polyacetylene},\ }\href
  {https://doi.org/10.1103/PhysRevLett.42.1698} {\bibfield  {journal} {\bibinfo
   {journal} {Phys. Rev. Lett.}\ }\textbf {\bibinfo {volume} {42}},\ \bibinfo
  {pages} {1698} (\bibinfo {year} {1979})}\BibitemShut {NoStop}%
\bibitem [{\citenamefont {Vladimirova}\ \emph {et~al.}(1998)\citenamefont
  {Vladimirova}, \citenamefont {Ivchenko},\ and\ \citenamefont
  {Kavokin}}]{Vladimirova1998}%
  \BibitemOpen
  \bibfield  {author} {\bibinfo {author} {\bibfnamefont {M.~R.}\ \bibnamefont
  {Vladimirova}}, \bibinfo {author} {\bibfnamefont {E.~L.}\ \bibnamefont
  {Ivchenko}},\ and\ \bibinfo {author} {\bibfnamefont {A.~V.}\ \bibnamefont
  {Kavokin}},\ }\bibfield  {title} {\bibinfo {title} {Exciton polaritons in
  long-period quantum-well structures},\ }\href
  {https://doi.org/10.1134/1.1187364} {\bibfield  {journal} {\bibinfo
  {journal} {Semiconductors}\ }\textbf {\bibinfo {volume} {32}},\ \bibinfo
  {pages} {90} (\bibinfo {year} {1998})}\BibitemShut {NoStop}%
\bibitem [{\citenamefont {Solnyshkov}\ \emph {et~al.}(2016)\citenamefont
  {Solnyshkov}, \citenamefont {Nalitov},\ and\ \citenamefont
  {Malpuech}}]{PhysRevLett.116.046402}%
  \BibitemOpen
  \bibfield  {author} {\bibinfo {author} {\bibfnamefont {D.~D.}\ \bibnamefont
  {Solnyshkov}}, \bibinfo {author} {\bibfnamefont {A.~V.}\ \bibnamefont
  {Nalitov}},\ and\ \bibinfo {author} {\bibfnamefont {G.}~\bibnamefont
  {Malpuech}},\ }\bibfield  {title} {\bibinfo {title} {Kibble-{Z}urek mechanism
  in topologically nontrivial zigzag chains of polariton micropillars},\ }\href
  {https://doi.org/10.1103/PhysRevLett.116.046402} {\bibfield  {journal}
  {\bibinfo  {journal} {Phys. Rev. Lett.}\ }\textbf {\bibinfo {volume} {116}},\
  \bibinfo {pages} {046402} (\bibinfo {year} {2016})}\BibitemShut {NoStop}%
\bibitem [{\citenamefont {St-Jean}\ \emph {et~al.}(2017)\citenamefont
  {St-Jean}, \citenamefont {Goblot}, \citenamefont {Galopin}, \citenamefont
  {Lema{\^\i}tre}, \citenamefont {Ozawa}, \citenamefont {Le~Gratiet},
  \citenamefont {Sagnes}, \citenamefont {Bloch},\ and\ \citenamefont
  {Amo}}]{st2017lasing}%
  \BibitemOpen
  \bibfield  {author} {\bibinfo {author} {\bibfnamefont {P.}~\bibnamefont
  {St-Jean}}, \bibinfo {author} {\bibfnamefont {V.}~\bibnamefont {Goblot}},
  \bibinfo {author} {\bibfnamefont {E.}~\bibnamefont {Galopin}}, \bibinfo
  {author} {\bibfnamefont {A.}~\bibnamefont {Lema{\^\i}tre}}, \bibinfo {author}
  {\bibfnamefont {T.}~\bibnamefont {Ozawa}}, \bibinfo {author} {\bibfnamefont
  {L.}~\bibnamefont {Le~Gratiet}}, \bibinfo {author} {\bibfnamefont
  {I.}~\bibnamefont {Sagnes}}, \bibinfo {author} {\bibfnamefont
  {J.}~\bibnamefont {Bloch}},\ and\ \bibinfo {author} {\bibfnamefont
  {A.}~\bibnamefont {Amo}},\ }\bibfield  {title} {\bibinfo {title} {Lasing in
  topological edge states of a one-dimensional lattice},\ }\href@noop {}
  {\bibfield  {journal} {\bibinfo  {journal} {Nat. Photonics}\ }\textbf
  {\bibinfo {volume} {11}},\ \bibinfo {pages} {651} (\bibinfo {year}
  {2017})}\BibitemShut {NoStop}%
\bibitem [{\citenamefont {Harrison}(1989)}]{harrison2012electronic}%
  \BibitemOpen
  \bibfield  {author} {\bibinfo {author} {\bibfnamefont {W.~A.}\ \bibnamefont
  {Harrison}},\ }\href@noop {} {\emph {\bibinfo {title} {Electronic structure
  and the properties of solids: the physics of the chemical bond}}}\ (\bibinfo
  {publisher} {Dover Publications, New York},\ \bibinfo {year} {1989})\ p.\
  \bibinfo {pages} {586}\BibitemShut {NoStop}%
\bibitem [{\citenamefont {Buchs}\ \emph {et~al.}(2021)\citenamefont {Buchs},
  \citenamefont {Marganska}, \citenamefont {Gonz\'{a}lez}, \citenamefont
  {Eimre}, \citenamefont {Pignedoli}, \citenamefont {Passerone}, \citenamefont
  {Ayuela}, \citenamefont {Gr\"{o}ning},\ and\ \citenamefont
  {Bercioux}}]{buchs2021metallic}%
  \BibitemOpen
  \bibfield  {author} {\bibinfo {author} {\bibfnamefont {G.}~\bibnamefont
  {Buchs}}, \bibinfo {author} {\bibfnamefont {M.}~\bibnamefont {Marganska}},
  \bibinfo {author} {\bibfnamefont {J.~W.}\ \bibnamefont {Gonz\'{a}lez}},
  \bibinfo {author} {\bibfnamefont {K.}~\bibnamefont {Eimre}}, \bibinfo
  {author} {\bibfnamefont {C.~A.}\ \bibnamefont {Pignedoli}}, \bibinfo {author}
  {\bibfnamefont {D.}~\bibnamefont {Passerone}}, \bibinfo {author}
  {\bibfnamefont {A.}~\bibnamefont {Ayuela}}, \bibinfo {author} {\bibfnamefont
  {O.}~\bibnamefont {Gr\"{o}ning}},\ and\ \bibinfo {author} {\bibfnamefont
  {D.}~\bibnamefont {Bercioux}},\ }\bibfield  {title} {\bibinfo {title}
  {Metallic carbon nanotube quantum dots with broken symmetries as a platform
  for tunable terahertz detection},\ }\href@noop {} {\bibfield  {journal}
  {\bibinfo  {journal} {Appl. Phys. Rev}\ }\textbf {\bibinfo {volume} {8}},\
  \bibinfo {pages} {021406} (\bibinfo {year} {2021})}\BibitemShut {NoStop}%
\bibitem [{\citenamefont {Valadbeigi}\ \emph {et~al.}(2016)\citenamefont
  {Valadbeigi}, \citenamefont {Ilbeigi},\ and\ \citenamefont
  {Farrokhpour}}]{valadbeigi2016ionization}%
  \BibitemOpen
  \bibfield  {author} {\bibinfo {author} {\bibfnamefont {Y.}~\bibnamefont
  {Valadbeigi}}, \bibinfo {author} {\bibfnamefont {V.}~\bibnamefont
  {Ilbeigi}},\ and\ \bibinfo {author} {\bibfnamefont {H.}~\bibnamefont
  {Farrokhpour}},\ }\bibfield  {title} {\bibinfo {title} {Ionization energies,
  electron affinities, and binding energies of {L}i-doped gold nanoclusters},\
  }\href@noop {} {\bibfield  {journal} {\bibinfo  {journal} {Res. Chem.
  Intermed.}\ }\textbf {\bibinfo {volume} {42}},\ \bibinfo {pages} {4921}
  (\bibinfo {year} {2016})}\BibitemShut {NoStop}%
\bibitem [{\citenamefont {Kaiser}\ \emph {et~al.}(2010)\citenamefont {Kaiser},
  \citenamefont {Sun}, \citenamefont {Lin}, \citenamefont {Chang},
  \citenamefont {Mebel}, \citenamefont {Kostko},\ and\ \citenamefont
  {Ahmed}}]{kaiser2010experimental}%
  \BibitemOpen
  \bibfield  {author} {\bibinfo {author} {\bibfnamefont {R.~I.}\ \bibnamefont
  {Kaiser}}, \bibinfo {author} {\bibfnamefont {B.~J.}\ \bibnamefont {Sun}},
  \bibinfo {author} {\bibfnamefont {H.~M.}\ \bibnamefont {Lin}}, \bibinfo
  {author} {\bibfnamefont {A.~H.}\ \bibnamefont {Chang}}, \bibinfo {author}
  {\bibfnamefont {A.~M.}\ \bibnamefont {Mebel}}, \bibinfo {author}
  {\bibfnamefont {O.}~\bibnamefont {Kostko}},\ and\ \bibinfo {author}
  {\bibfnamefont {M.}~\bibnamefont {Ahmed}},\ }\bibfield  {title} {\bibinfo
  {title} {An experimental and theoretical study on the ionization energies of
  polyynes ({H}-({C}$\equiv$ {C}) n-{H}; n= 1-9)},\ }\href@noop {} {\bibfield
  {journal} {\bibinfo  {journal} {Astrophys. J}\ }\textbf {\bibinfo {volume}
  {719}},\ \bibinfo {pages} {1884} (\bibinfo {year} {2010})}\BibitemShut
  {NoStop}%
\bibitem [{\citenamefont {Davies}\ \emph {et~al.}(2008)\citenamefont {Davies},
  \citenamefont {Burnett}, \citenamefont {Fan}, \citenamefont {Linfield},\ and\
  \citenamefont {Cunningham}}]{davies2008terahertz}%
  \BibitemOpen
  \bibfield  {author} {\bibinfo {author} {\bibfnamefont {A.~G.}\ \bibnamefont
  {Davies}}, \bibinfo {author} {\bibfnamefont {A.~D.}\ \bibnamefont {Burnett}},
  \bibinfo {author} {\bibfnamefont {W.}~\bibnamefont {Fan}}, \bibinfo {author}
  {\bibfnamefont {E.~H.}\ \bibnamefont {Linfield}},\ and\ \bibinfo {author}
  {\bibfnamefont {J.~E.}\ \bibnamefont {Cunningham}},\ }\bibfield  {title}
  {\bibinfo {title} {Terahertz spectroscopy of explosives and drugs},\
  }\href@noop {} {\bibfield  {journal} {\bibinfo  {journal} {Mater. Today}\
  }\textbf {\bibinfo {volume} {11}},\ \bibinfo {pages} {18} (\bibinfo {year}
  {2008})}\BibitemShut {NoStop}%
\bibitem [{\citenamefont {Hartmann}\ \emph {et~al.}(2019)\citenamefont
  {Hartmann}, \citenamefont {Saroka},\ and\ \citenamefont
  {Portnoi}}]{hartmann2019interband}%
  \BibitemOpen
  \bibfield  {author} {\bibinfo {author} {\bibfnamefont {R.~R.}\ \bibnamefont
  {Hartmann}}, \bibinfo {author} {\bibfnamefont {V.~A.}\ \bibnamefont
  {Saroka}},\ and\ \bibinfo {author} {\bibfnamefont {M.~E.}\ \bibnamefont
  {Portnoi}},\ }\bibfield  {title} {\bibinfo {title} {Interband transitions in
  narrow-gap carbon nanotubes and graphene nanoribbons},\ }\href@noop {}
  {\bibfield  {journal} {\bibinfo  {journal} {J. Appl. Phys}\ }\textbf
  {\bibinfo {volume} {125}},\ \bibinfo {pages} {151607} (\bibinfo {year}
  {2019})}\BibitemShut {NoStop}%
\bibitem [{\citenamefont {Hu}\ and\ \citenamefont
  {Feng}(1991)}]{hu1991feasibility}%
  \BibitemOpen
  \bibfield  {author} {\bibinfo {author} {\bibfnamefont {Q.}~\bibnamefont
  {Hu}}\ and\ \bibinfo {author} {\bibfnamefont {S.}~\bibnamefont {Feng}},\
  }\bibfield  {title} {\bibinfo {title} {Feasibility of far-infrared lasers
  using multiple semiconductor quantum wells},\ }\href@noop {} {\bibfield
  {journal} {\bibinfo  {journal} {Appl. Phys. Lett}\ }\textbf {\bibinfo
  {volume} {59}},\ \bibinfo {pages} {2923} (\bibinfo {year}
  {1991})}\BibitemShut {NoStop}%
\end{thebibliography}%

\end{document}